\documentclass[a4paper]{quantumarticle}
\pdfoutput=1
\usepackage[utf8]{inputenc}
\usepackage[english]{babel}
\usepackage[T1]{fontenc}
\usepackage{amsmath}
\usepackage{amssymb}
\usepackage{hyperref}
\usepackage{graphicx}
\usepackage{array}
\usepackage{multirow}
\usepackage{tikz}
\usepackage{lipsum}
\usepackage{booktabs}

\begin{document}

\title{Non-Relativistic Quantum Mechanics
in Multidimensional Geometric Frameworks}

\author{Dalaver H. Anjum*}
\affiliation{Department of Physics, Khalifa University, P.O. Box 127788, Abu Dhabi, United Arab Emirates}
\orcid{0000-0003-2336-2859}
\email{dalaver.anjum@ku.ac.ae}
\author{Shahid Nawaz}
\affiliation{Shaker High School, 445 Watervliet Shaker Rd, Latham, NY 12110, United States of America}
\author{Muhammad Saleem}
\affiliation{Department of Physics, Bellarmine University, 2001 Newburg Road, Louisville, KY 40205, United States of America}

\begin{abstract}\label{abs}
\raggedright
We investigate a generalized formulation of non-relativistic quantum mechanics associated with multidimensional geometric frameworks in which the dispersion relation takes the power-law form \(E \propto |p|^{j}\), with \(j=N-1\). Quantization of this modified kinetic structure leads to a \(j\)-th order Schr\"odinger-type equation, which we study for free particles and for a particle confined in a one-dimensional infinite potential well. The formalism is worked out explicitly for the 2G, 3G, 4G, and 5G cases. In the 3G limit, the standard quadratic dispersion relation, sinusoidal bound-state wavefunctions, and familiar energy spectrum are recovered. By contrast, the 4G and 5G cases give rise to third- and fourth-order boundary-value problems whose admissible eigenfunctions are composed of geometry-dependent combinations of exponential, trigonometric, and hyperbolic terms, with corresponding spectra scaling as \(E_n \propto n^3\) and \(E_n \propto n^4\), respectively. In the 2G case, the linear kinetic structure does not support nontrivial bound states under the Dirichlet boundary conditions considered here. We further formulate generalized definitions of probability density, expectation values, and commutator relations adapted to this setting, and examine the associated uncertainty relation within the proposed framework. Taken together, these results provide a concrete characterization of how higher-order geometric kinetic operators modify free-particle dispersion, confinement, and spectral structure relative to conventional non-relativistic quantum mechanics.
\end{abstract}

Quantum mechanics is one of the most successful theories of the 20th century~\cite{Ballentine2014, CohenTannoudji1991}. It is a fundamental theory of physics that describes the behavior of matter and energy at the atomic and subatomic scales. It greatly enhanced our understanding of the universe and is the backbone of modern technology, from transistors in computers to lasers in medical devices. Quantum mechanics, in both nonrelativistic and relativistic limits, has enabled significant progress in understanding fundamental and applied physics. The non-relativistic quantum mechanics (NRQM) is formulated in the Hilbert space $\mathbb{H}^{n}$ ~\cite{ReedSimon1980, Hall2013}, which is an n-dimensional complex vector space. 

Quantum mechanics is built on several fundamental principles. The first is that physical systems are described by wavefunctions or, more generally, state vectors in a Hilbert space. This framework unifies the apparently dual wave-like and particle-like aspects of matter: the state evolves as a probability amplitude capable of interference and superposition, while measurement outcomes are registered as localized quanta. Thus, wave-particle duality is not a coexistence of two classical pictures, but a manifestation of the quantum state’s dual role in propagation and measurement. A second central principle is quantization, whereby observable quantities are represented by linear operators whose spectra determine the allowed measurement outcomes. In bound systems, this often leads to discrete eigenvalue spectra, as in the quantized energy levels of electrons in atoms or the quantized angular momentum of confined states. Finally, quantum mechanics is fundamentally probabilistic. The state of a system does not, in general, determine unique outcomes for individual measurements, but instead specifies probabilities for the different possible results. These probabilities are encoded through the Born rule and reflect the operator structure of the theory, especially the fact that noncommuting observables cannot generally be measured simultaneously with arbitrary precision. This gives rise to the uncertainty principle, which places a fundamental bound on the joint sharpness of observables such as position and momentum~\cite{Heisenberg1927, Robertson1929, Schrodinger1930}, which states a fundamental limit to the precision with which specific pairs of properties can be known simultaneously. The state of a system is described by a wavefunction, or more generally by a state vector in Hilbert space. This state encodes the full probabilistic content of the theory, in the sense that it determines the probability amplitudes and expectation values associated with all measurable observables. A defining feature of the formalism is the superposition principle: whenever two state vectors are allowed, any linear combination of them is also a valid physical state. As a result, a quantum system may exist in a coherent superposition of distinct configurations, giving rise to interference effects that have no classical analog. Upon measurement, one obtains a single outcome drawn according to the Born probabilities associated with the state, and the state is correspondingly updated by the measurement process. In Schrödinger’s formulation, the fundamental concepts outlined above are encoded in a single differential equation governing the time evolution of the quantum state \cite{Schrodinger1926, GriffithsSchroeter2018,vonNeumann1955, Dirac1958}: 
\begin{equation}
 \hat{H}_{3G} \psi_{3G} = \hat{E}_{3G} \psi_{3G}
 \label{eq:1}
\end{equation} 

Where $ \psi_{3G} $ represents the wave function of quantum mechanical particles, $\hat{H}_{3G}= \hat{K.E.}_{3G} + \hat{P.E.}_{3G}$ is the Hamiltonian operator representing the particle's energy operator,  $ \hbar $ is the reduced Planck constant, and $t$ is time. The kinetic energy ($K.E.$) of a particle with mass $m$ and linear momentum $p$ is equal to $\frac{p^2}{2m}$, which is a consequence of the work-K.E. theorem. In addition, the non-relativistic limit of the particle's relativistic energy ($E=\gamma_{3G}mc^2$) is given by $K.E.=(\gamma_{3G}-1)mc^2$ \cite{Einstein1905Inertia,Griffiths2017Electrodynamics,das1993}.  The $\gamma_{3G}=(1-\frac{v^2}{c^2})^{-1/2}$ is the relativistic Lorentz factor that comes from the Lorentz transformation to describe the spacetime relativity between the inertial frames. If the moving frame's speed, $v$, is low compared to the speed of light, the $K.E.$ reduces to $\ \frac {p^2_{3G}}{2m}$. It is evidence that the nonrelativistic form of $K.E.$ is also a consequence of the Minkowski distance form in the framework. A particle that moves in a potential $V_{3G}(x_{1}, x_{2}, ...x_{n}, t)$ in a vector space of $n$ dimensions $\hat{H}$ becomes equal to the sum of its $K.E.$, and $P.E.$. The corresponding form of the S.E. of Eq.~\ref{eq:1} can then be written in the alternative form:
\begin{widetext}
\begin{equation}
\left(
\frac{1}{2m} \sum_{i=1}^{n} \hat{p}^{\,2}_{i3G}
+ \hat{V}_{3G}(x_1, x_2, \ldots, x_n, t)
\right)
\psi_{3G}(x_1, x_2, \ldots, x_n, t)
=
\hat{E(t)}_{3G}\,\psi_{3G}(x_1, x_2, \ldots, x_n, t)
\label{eq:2}
\end{equation}
\end{widetext}
Where the $\hat{p_{i3G}}$, $\hat{V}_{3G}(x_1, x_2, \ldots, x_n, t)$, and $\hat{E(t)}_{3G}$ are linear momentum, potential energy, and total energy operators in Eq.~\ref{eq:2} in the 3G framework. This formulation of NRQM is for the n-dimensional Hilbert vector space $\mathbb{H}^{n}$ \cite{GriffithsSchroeter2018}. \\
Recent developments in generalized quantum dynamics have explored extensions of the standard Schr\"odinger equation beyond the conventional quadratic kinetic operator. A notable example is the fractional Schr\"odinger equation introduced by Laskin, in which the kinetic term is defined through a nonlocal fractional Laplacian, leading to a dispersion relation of the form \( E \propto |p|^{\alpha} \) with \( 0 < \alpha \leq 2 \) \cite{Laskin2000, Laskin2002, Herrmann2014}. This framework arises from path integrals over L\'evy flights and captures anomalous transport and nonlocal quantum behavior. In parallel, higher-order partial differential equation (PDE) models have been investigated, in which the kinetic energy operator involves integer powers of momentum, leading to higher-order spatial derivatives and modified spectral properties \cite{KleinGordon1926, BenderBoettcher1998, AblowitzMusslimani2013}. Such models have appeared in effective theories, generalized wave equations, and studies of dispersive media. Additionally, non-Hermitian and multi-phase formulations of quantum mechanics have introduced extended complex structures, including systems with multiple conjugate components or nontrivial phase algebras, often motivated by PT-symmetry or generalized inner-product spaces \cite{Bender2007, Mostafazadeh2002a, Mostafazadeh2002b, Longhi2015}.\\
Despite these advances, most existing approaches introduce modified kinetic operators either phenomenologically or through stochastic or algebraic generalizations, without a direct connection to an underlying geometric principle. In contrast, the present work develops a geometry-driven formulation in which the structure of the quantum operator is derived from a generalized Minkowski distance defined on \( L^{j} \)-normed spaces \cite{xu2018evolutionary, materum2021improved, merigo2011new}. Within this framework, the dispersion relation and the corresponding \( j \)-th order Schr\"odinger equation emerge naturally from the geometric properties of the underlying space, rather than being imposed externally. In this way, a direct link between the geometry of the configuration space and the dynamical laws governing quantum evolution has been attempted to be established, providing a unified and systematic extension of non-relativistic quantum mechanics to higher-dimensional geometric settings. It is started from using Minkowski's distance (MD) formulation for measuring distance in N-dimensional geometry, i.e., if $A(x_i,x_2,\ldots, x_n)$ and $B(x^\prime_1, x_2^\prime \ldots, x_n^\prime)$ are two points in an n-dimensional vector space $\mathbb{R}^n$, then the MD ($\Delta s_{NG}$) between these points can be written in the following way \cite{xu2018evolutionary, materum2021improved, merigo2011new, Minkowski1896}:
\begin{equation} 
    \Delta s_{NG}=\left(\sum_{i=1}^{n}(x_i^\prime-x_i)^j\right)^{1/j}\label{Minkowski-1}
\end{equation}
Where $j=N-1$. For the 3G case, $\Delta s_{NG}$ becomes $\Delta s_{3G}$, which forms the basis of any physics that involves measuring distances, including classical physics, electromagnetism, and quantum mechanics.  The question remains in the case of N-dimensional geometry (NG), which can be understood as quantum mechanics in $L^j(\mathbb{R}^{n})$ space \cite{ReedSimon1972,Rudin1987, Triebel1983}. In other words, how the NRQM will look in different dimensional geometries, including 2G, 4G, and beyond. This requires using the non-relativistic $K.E.$ of a particle with mass $m$ and speed $v$ in the NG framework, which can be written in the following form \cite{anjum2026}:

\begin{equation}
(K.E.)_{non-rel.}=\frac{1}{j} \left(\frac{v}{c}\right)^{j} mc^2= \frac{1}{j} \frac{p^j}{m^{j-1}c^{j-2}}\label{gamma-k}
\end{equation} 

The generalized non-relativistic kinetic energy expressed in Eq.~\ref{gamma-k} naturally raises the question of how the fundamental dynamical equation of quantum mechanics should be formulated within the NG framework. In the standard 3G case, the Schr\"odinger equation follows from promoting the classical kinetic energy $p^{2}/2m$ to an operator acting on a wavefunction in Hilbert space. By analogy, the modified dispersion relation implied by Eq.~\ref{gamma-k} suggests that a corresponding generalized Schr\"odinger equation must be constructed in which the kinetic operator involves $j$-th order spatial derivatives. Such an equation is required to consistently describe the quantum dynamics of particles when the underlying geometric structure modifies the momentum–energy relation.

In the conventional formulation of quantum mechanics, the functional form of the potential energy $V _ {3G} (\mathbf{r}) $ is typically determined by well-established physical principles, such as symmetry, conservation laws, and field equations. For example, the Coulomb potential follows from Gauss's law in three spatial dimensions, while the harmonic oscillator potential emerges from linear restoring forces. In the NG framework, however, the geometric structure differs from the standard Euclidean case, and deriving interaction laws is no longer straightforward. The precise form of $V(\mathbf{r})$ consistent with NG geometry would, in principle, require a reformulation of classical field theory and force laws within that same geometric setting. Consequently, the exact functional dependence of physically realistic potentials in NG frameworks cannot be uniquely specified at this stage. For this reason, it is both natural and necessary to first investigate quantum systems in which the potential energy takes a simplified form. Two particularly important cases are those in which $V(\mathbf{r})_{NG} = 0$, corresponding to free particles described by plane-wave solutions, and those in which the particle is confined within a finite spatial region by an infinite potential barrier. The latter corresponds to a particle in a box with $V_{NG} = 0$ inside the domain and $V_{NG} = \infty$ outside. These systems do not rely on a detailed specification of the interaction potential but instead impose boundary conditions that isolate the effects of the generalized kinetic operator. By studying free and confined particles in an infinite potential well, the essential features of quantization, spectral scaling, and wavefunction structure in the NG framework can be examined without introducing additional ambiguities arising from unknown interaction laws.\\
Motivated by the geometric origin of dynamical laws, we develop a generalized formulation of non-relativistic quantum mechanics within $N$-dimensional geometric (NG) frameworks characterized by a modified power-law dispersion relation $E \propto |p|^{j}$ with $j = N-1$. By extending the standard quadratic kinetic operator to higher-order spatial derivatives, a corresponding generalized Schr\"odinger equation is constructed and applied to representative quantum systems. The formalism is systematically implemented for 2G, 3G, 4G, and 5G cases, and explicit solutions are obtained for free particles and for particles confined within an infinite potential well. The resulting eigenfunctions, eigenenergies, canonical bracket relations, and generalized uncertainty products are analyzed and compared across geometries, highlighting how departures from quadratic dispersion modify spectral scaling, wavefunction structure, and statistical properties. This work thus establishes a consistent framework for exploring quantum dynamics beyond the conventional Laplacian structure of standard non-relativistic quantum mechanics.

\section{Formulation of the NG Framework for Non-Relativistic Quantum Mechanics}

In 3G framework, as stated earlier that the functional form of the potential $V_{3G}(r)$ is typically guided by well-established physical principles, symmetry arguments, and experimental observations. For example, the Coulomb potential arises from Gauss’s law in three spatial dimensions, and the harmonic oscillator potential emerges naturally from quadratic restoring forces in classical mechanics. However, in the NG framework with $j \neq 2$, the underlying geometric structure differs from the Euclidean 3G case, and the standard derivations of physically motivated potentials no longer apply straightforwardly. In particular, the form of central forces derived from flux conservation arguments depends explicitly on spatial dimensionality, and therefore extending familiar potentials such as $1/r$ or $r^2$ to higher-order geometric frameworks is not unique or unambiguous. Moreover, the modified dispersion relation $E_{NG} \propto |p|^{j}$ changes the balance between kinetic and potential contributions in the Hamiltonian. As a result, even if one were to postulate a functional form for $V_{NG}(r)$ by analogy with the 3G case, its physical interpretation and scaling behavior would generally differ. Without a fully developed field-theoretic or symmetry-based derivation of interaction laws in NG geometries, determining the exact and physically consistent form of $V_{NG}(r)$ becomes highly nontrivial.\\
For this reason, the present work focuses on a quantum system in which the potential is specified as simply as possible: $V_{NG}(x)=0$ within a finite spatial region and $V_{NG}(x)=\infty$ outside that region. This corresponds to a particle confined in a one-dimensional box of infinite potential height. Consequently, the particle in an infinite potential well does not rely on a detailed functional dependence of $V_{NG}(r)$. Nevertheless, this imposes boundary conditions that confine the particle to a finite domain, therefore, providing a framework for analyzing how the modified kinetic operator in the NG formulation affects quantization, eigenfunctions, and spectral scaling in higher-dimensional geometric settings.\\
The confined particle in an infinite potential box serves as a minimal and robust test case for exploring the consequences of NG quantum mechanics. It isolates the effects of the generalized kinetic term while avoiding ambiguities associated with constructing physically motivated potentials beyond the standard 3G framework. The non-relativistic $K.E.$ of a particle takes the form of $pc$, $\frac{p^2}{2m}$, $\frac{p^3}{3m^2c}$, and $\frac{p^4}{4m^{3}c^{2}}$ for 2G, 3G, 4G, and 5G, respectively. The particle in the one-dimensional potential well of width $l$ and infinite height $(V_{NG}=\infty)$ can be represented by a schematic in Figure 1. The SE for a quantum particle that is in a potential $V$ in the NG framework can be expressed in the following form:
\begin{widetext}
\begin{equation}
\left(
\frac{1}{j}\,\frac{\hat{p}_{\mathrm{NG}}^{\,j}}{m^{\,j-1}c^{\,j-2}}
+\hat{V}_{\mathrm{NG}}(x_{1},x_{2},\ldots,x_{n},t)
\right)
\psi_{\mathrm{NG}}(x_{1},x_{2},\ldots,x_{n},t)
=
\hat{E}_{\mathrm{NG}}\,
\psi_{\mathrm{NG}}(x_{1},x_{2},\ldots,x_{n},t)
\end{equation}
\end{widetext}      
The operator forms of the $\hat{p_{x}}$, $\hat{V}(x_1, x_2, \ldots, x_n, t)$, and $\hat{E(t)}$ operators in the NG framework must be writen down.  In particular, the exact form of the $\hat{V}(x_1, x_2, \ldots, x_n, t)$ operator will be the most difficult to determine, as it depends on the system's configuration. The usual 3G framework related $\hat{p_{x}}$, and $\hat{E(t)}$ operators of forms $\hat{p_{x}} = -\iota \hbar \frac{\partial}{\partial x}=(-1) (-1)^{\frac{1}{3-1}} \hbar \frac{\partial}{\partial x}$
and, $\hat{E} = i\hbar \frac{\partial}{\partial t}=(+1) (-1)^{\frac{1}{3-1}} \hbar \frac{\partial}{\partial t}$ can be utilized to define the generalized forms of these operators for the NG frameworks. In quantum physics, energy and momentum are not merely measured quantities; they are the generators of change in time and space \cite{Shankar1994,SakuraiNapolitano2017}. Momentum is the operator associated with spatial translation of the form $\hat{T}(\Delta x)=\exp{\left(\frac{\iota}{\hbar}\hat{p}\Delta x\right)}$: when a wavefunction is shifted from one position to another, the transformation is generated by the momentum operator, reflecting the deep fact that momentum governs how quantum states respond to spatial displacement. Likewise, energy plays the analogous role for time evolution of the form $\hat{U}(\Delta t)=\exp{\left(-\frac{\iota}{\hbar}\hat{E}\Delta t\right)}$: it is the generator of translations forward or backward in time, determining how a wavefunction changes as the system evolves. This is why the momentum operator appears as a spatial derivative and the energy operator as a time derivative: derivatives measure infinitesimal change, and generators encode the structure of continuous translations. In the 3G framework, the momentum generates spatial translations, so shifting a wavefunction in space produces a phase factor with the momentum term, while energy generates time translations, so evolution in time produces a phase factor with the energy term. The opposite signs arise from the plane-wave phase \(e^{\frac{i}{\hbar}(px-Et)}\), where space enters as \(+px\) and time enters as \(-Et\). At a deeper level, this reflects the symmetry of nature itself---homogeneity of space gives rise to momentum conservation, while homogeneity of time gives rise to energy conservation---so in quantum mechanics, energy and momentum are the mathematical embodiments of the invariance of physical law under translations in time and space. Clearly, in the 3G framework, one can envisage that the square roots of negative unity serve as a mathematical operator in the complex wave function $\Psi_{3G}(x,t) = Ae^{i(kx - \omega t)}$, where it encodes the phase $\phi = kx - \omega t$ into a complex exponent. This allows the wave's oscillation to be represented as a rotating vector in the complex plane, simplifying the calculation of interference and diffraction.
 In this spirit, in the NG frameworks, we define the $\hat{p}$, and $\hat{E}$ operator forms by using the roots of negative unity in the following way:
\begin{widetext}
\begin{equation}
\begin{cases}
\hat{p} = +\left[j^{th}~ root~of~(-1)^{1/j}\right] \hbar \frac{\partial}{\partial x}, 
\quad
\hat{E} = +\left[j^{th} root ~of~(-1)^{1/j}\right] \hbar \frac{\partial}{\partial t}, 
& j=1 
\\[8pt]
\hat{p} = -\left[j^{th}~ root ~of~(-1)^{1/j}\right] \hbar \frac{\partial}{\partial x}, 
\quad
\hat{E} = +\left[j^{th} root ~of~(-1)^{1/j}\right] \hbar \frac{\partial}{\partial t}, 
& j\ge2
\end{cases}
\label{eq:6}
\end{equation}
\end{widetext}

In Eq.~\ref{eq:6}, the $j^{th}$ roots $\left(a_{j}\right)$ of negative unity ($(-1)^{1/j}$) are utilized, which can be determined by expressing $(-1)^{1/j}$ in the complex polar form, i.e., $z^{1/j}=r^{1/j} e^{\iota \pi\left(\frac{2l+1}{j} \right)}$, with $l=0, 1, 2, ....j-1$. The $\iota$ and $-\iota$ are the square roots of $-1$, of course. And the cube roots of $-1$ are $\bar{1}$, $\bar{\omega}$, and $\bar{\omega}^{2}$ in which $\bar{1}=-1$, $\bar{\omega}=\frac{1+\iota \sqrt{3}}{2}$, and $\bar{\omega}^{2}=\frac{1-\iota \sqrt{3}}{2}$. Similarly, the quartic roots of the negative unity are four roots, namely $\bar{\eta_{1}}=\frac{1+\iota}{\sqrt {2}}$, $\bar{\eta_{2}}=\frac{1-\iota}{\sqrt {2}}$, $\bar{\eta_{3}}=\frac{-1+\iota}{\sqrt {2}}$, and $\bar{\eta_{4}}=\frac{-1-\iota}{\sqrt {2}}$. Under this scheme, the forms of linear momentum operators say along the x-axis, $\hat{p}$, become $\frac{\partial}{\partial x}$, $-\iota \frac{\partial}{\partial x}$, $\bar{1}\bar{\omega_{1}} \frac{\partial}{\partial x}$, and $\bar{\eta}_1 \bar{\eta}_2 \bar{\eta}_3 \frac{\partial}{\partial x}$ for 2G, 3G, 4G and 5G frameworks, respectively. Since the product of all \(j\) roots equals $(-1)^j(-1) = (-1)^{j+1}$
up to the monic polynomial sign rule, it follows in particular that the product of any \(j-1\) of these roots is equal to the reciprocal of the remaining root. Because each root lies on the unit circle, this reciprocal is the complex conjugate of the omitted \(j\)-th root. For instance, the product $\bar{\eta}_1 \bar{\eta}_2 \bar{\eta}_3$ can also be written as: $-\bar{\eta}_4$. The operator forms for energy $(\hat{E})$ contain a positive $+$ prefactor to the $j^{th}$ root, therefore these are written as $\frac{\partial}{\partial t}$, $\iota \frac{\partial}{\partial t}$, $\bar{\omega}^{2} \frac{\partial}{\partial t}$, and $\bar{\eta_{4}} \frac{\partial}{\partial t}$ for 2G, 3G, 4G, and 5G frameworks, respectively. Using the roots of negative unity under this scheme, as an example, the operator forms for $p$ and $E$ for $j= 1 to 4$ are given below:
\begin{equation}
\begin{cases}
\hat{p} = -\hbar\dfrac{\partial}{\partial x}, 
\quad
\hat{E} = -\hbar\dfrac{\partial}{\partial t}, 
& j=1 
\\[8pt]
\hat{p} = -\iota \hbar\dfrac{\partial}{\partial x}, 
\quad
\hat{E} = +\iota \hbar\dfrac{\partial}{\partial t}, 
& j=2
\\[8pt]
\hat{p} = -\bar{\omega}^{2} \hbar\dfrac{\partial}{\partial x}, 
\quad
\hat{E} = \bar{\omega}^{2} \hbar\dfrac{\partial}{\partial t}, 
& j=3 \\[8pt]
\hat{p} = -\bar{\eta}_{4}\,\hbar\dfrac{\partial}{\partial x}, 
\quad
\hat{E} = \bar{\eta}_{4}\,\hbar\dfrac{\partial}{\partial t}, 
& j=4
\end{cases}
\label{eq:7}
\end{equation}

With these operator definitions, the Schrödinger equation for a particle subject to a potential 
\(V_{NG}(x_{1}, x_{2}, \dots, x_{n}, t)\) in an \(n\)-dimensional vector space within the NG framework 
can be expressed as a natural generalization of the standard quantum mechanical form. In this formulation, 
the spatial derivatives are elevated to \(j^{\text{th}}\)-order operators, reflecting the underlying 
algebra of the \(j^{\text{th}}\) roots of negative unity, while the temporal evolution is governed by a 
corresponding root-dependent prefactor. The resulting equation incorporates both the modified kinetic 
structure and the generalized potential term, ensuring consistency with the extended operator algebra. 
Moreover, this construction preserves the essential linearity of the Schrödinger equation while embedding 
higher-order geometric and algebraic features intrinsic to the NG framework. The explicit form of this 
generalized equation is given by:
\begin{widetext}
\begin{equation}
 \left[- \frac{1}{j} \frac{\hbar^{j}}{m^{j-1}c^{j-2}} \left(\sum_{i=1}^{n} \frac{\partial^{j}}{\partial x_{i}^{j}}\right)+ \hat{V}_{NG}(x_{1}, x_{2},....x_{n},t)\right] \psi_{NG} (x_{1},x_{2}...x_{n},t) = (-1)^{1/j} \hbar \frac{\partial}{\partial t} \psi_{NG} (x_{1},x_{2}...x_{n},t)
 \label{eq:8}
\end{equation} 
\end{widetext}
The Eq.~\ref{eq:8} can be reduced to a free quantum particle ($\hat{V}_{NG}=0$) in an n-dimensional Hilbert ($\mathbb{H}^{n}$) vectorspace in the following way: 
\begin{widetext}
\begin{equation}
 - \frac{1}{j} \frac{\hbar^{j}}{m^{j-1}c^{j-2}} \left(\sum_{i=1}^{n} \frac{\partial^{j}}{\partial x_{i}^{j}}\right) \psi_{NG} (x_{1},x_{2}...x_{n},t) = (-1)^{1/j} \hbar \frac{\partial}{\partial t} \psi_{NG} (x_{1},x_{2}...x_{n},t)
 \label{eq:9}
\end{equation}
\end{widetext}

This equation can be further reduced to a one-dimensional free quantum particle in the following form: 

\begin{equation}
 - \frac{1}{j} \frac{\hbar^{j}}{m^{j-1}c^{j-2}} \frac{\partial^{j}}{\partial x^{j}} \psi_{NG} (x,t) = (-1)^{1/j} \hbar \frac{\partial}{\partial t} \psi_{NG} (x,t) 
 \label{eq:10}
\end{equation} 

Therefore, for even and odd values of $j$, the $-$ and $+$ signs appear on the left-hand side of the equation, respectively. The particle's motion in a potential well demonstrates the utility of quantum mechanics in the NG framework. For the case of even $j$, the eigenfunctions $\left(\psi_{NG,n}\right)$ and eigenenergies $\left(E_{NG,n}\right)$ of a particle in a one-dimensional potential box can be given below: 
\begin{widetext}
\begin{equation}
\psi_{NG,n}(x,t) = \left[ A f(kx) \right] 
\left[ \exp\left( (j-1)~roots~of~(-1)^{\frac{1}{j}} \frac{E_{NG,n}}{\hbar} t \right) \right]
=\left[ A f(kx) \right] 
\left[ \exp\left( -j^{th}~root~of~(-1)^{\frac{1}{j}} \frac{E_{NG,n}}{\hbar} t \right) \right]
\end{equation}
\end{widetext}
Where $A\equiv 1$ is an arbitrary constant for the spatial part $\left(\phi_{NG,n}(x)=A f(kx)\right)$ of the eigenfunction $\psi_{NG,n}(x, t)$.  The wavefunction $\phi_{NG,n}(x)$ obeys time-independent SE as given below in the following form:
\begin{equation}
-\frac{1}{j}\frac{\hbar^{j}}{m^{j-1}c^{j-2}} \frac{\partial^{j}}{\partial x^{j}} \phi_{NG,n} (x) = E_{NG,n} \phi_{NG,n} (x) 
\end{equation} 

with; 
\begin{equation}
\begin{cases}
\phi_{NG,n} (x)=\sum_{j=1}^{N-1} C_{j}\exp{\left(a_{j}k_{NG}x\right)}
\\[8pt]
E_{NG,n} = \frac{\hbar^{j} k_{NG,n}^{j}}{{j}m^{j-1}c^{j-2}}
\end{cases}
\end{equation}

In the NG framework with $j=N-1$, the natural state space is not a Hilbert space for $j\neq 2$, but the Banach space $L^j(\Omega;\mathbb{C})$ with norm $\|\psi\|_j=\left(\int_\Omega |\psi(x)|^j\,dx\right)^{1/j}$~\cite{Brezis2011, Bobrowski2024}. Physical pure states are represented by branch-complete $j$-tuples $\mathbf{\Psi}=(\Psi_1,\dots,\Psi_j)$, and the fundamental pairing is the continuous $j$-linear functional $\Lambda_j(\Psi_1,\dots,\Psi_j)=\int_\Omega \prod_{r=1}^j \Psi_r(x)\,dx$, which is well-defined by H\"older's inequality ~\cite{Brezis2011}. For branch-generated states of the form $\Psi_r(x,t)=R(x,t)e^{a_r S(x,t)}$, with $a_r^{\,j}=-1$, $\sum_{r=1}^j a_r=0$, $R\ge 0$, and $R(\cdot,t)\in L^j(\Omega)$, the probability density becomes
\begin{equation}
\rho_j(x,t)=\prod_{r=1}^j \Psi_r(x,t)=R(x,t)^j\ge 0,
\label{eq:density}
\end{equation}
so that, $\int_\Omega \rho_j(x,t)\,dx=1$ provides the normalization condition. Thus, the standard Hilbert-space structure of 3G is replaced, for $j\neq 2$, by an $L^j$-Banach-space formulation equipped with a geometry-adapted multilinear pairing.
As in the case of 3G, the wave function of a free quantum particle is complex function and is represented as $\psi_{1}(x, t)=C \exp{\left(\iota k_{3G}x\right)} \exp{\left(-\iota \frac{E_{3G}}{\hbar} t\right)}$, with $C$ as a normalization constant, $k_{3G}$ the wavenumber and energy $E_{3G}$. The multiplication of $\psi_{2} (x,t)$ and its complex conjugate $(\psi_{1}(x,t))$, obtained by replacing $\iota$ with $-\iota$ in the phase of $\psi_{2} (x, t)$, gives probability density  $(\rho_{3G})$ of finding the particle, provided $\psi_{3G} (x,t)$ are square $(L^2)$ integrable. The probability $(P_{3G})$ can be obtained by integrating the square of the $(\rho_{3G})$ over $x$ from $-\infty$ to $\infty$, which is mathematically expressed as $P_{3G}=\int_{-\infty}^{+\infty} \psi_{1}(x,t) \psi_{2} (x, t) dx$. Hence, in the 3G framework, there are only two square roots of negative unity, so each wave function has only one corresponding complex conjugate. We extend this notion herein by stating that the wavefunction of a quantum particle in the NG framework has $j-1$ complex conjugates, where $j=N-1$ is the number of roots of negative unity. Consequently, the corresponding probability densities $(\rho_{NG})$ of finding particles can be obtained by carrying out $j$-fold conjugation of the wavefunctions, which can be followed by integrating these densities to find the probabilities $(P_{NG})$ in the limit that the wavefunction is an $L^{j}$ integrable function \cite{HornJohnson2013, Rudin1991, Folland1999, Stein2011}. Therefore, we present a definition of $P_{NG}$ in the following way, which is dictated by $j$-fold conjugations and $L^{j}$-integrable wavefunctions, $\psi_{NG,n}(x,t)$:
\begin{equation}
P_{NG,n} = \int_{-\infty}^{+\infty} \psi_{1,n} \psi_{2,n} \psi_{3,n} \cdots \psi_{j-1,n} \psi_{j,n} \, dx.
\label{eq:14}
\end{equation}
The probability definition given in Eq.~\ref{eq:14} is constructed to ensure that the resulting quantity remains real-valued within the \(L^{j}\)-normed geometric framework. In contrast to the conventional \(L^{2}\) case, where a single complex conjugate guarantees reality through \(|\psi|^{2}\), the NG formulation incorporates all \(j\) roots of negative unity, leading to a multi-phase structure of the wavefunction. The product of the \(j\) conjugate components forms a symmetric combination in which the complex phases cancel pairwise (or cyclically), yielding a real-valued integrand. The expectation value of an operator $\hat{O}$ in the NG framework can be calculated by multiplying the $\hat{O}\psi_{j}$ with the $j-1$ complex conjugates, which is essentially a generalization of finding the expectation values of operators in the 3G framework with $L^{2}$ integrable functions. In analogy with the standard Hilbert-space formulation, the expectation value of a hermitian operator $\hat{O}$ is given by:
\begin{widetext}
\begin{equation}
\langle \hat{O} \rangle_{NG}
=
\frac{
\int_{-\infty}^{+\infty}
\psi_1 \psi_2 \psi_3 \cdots \psi_{j-1}
\, \hat{O}\,\psi_j \, dx
}{
\int_{-\infty}^{+\infty}
\psi_1 \psi_2 \psi_3 \cdots \psi_{j-1}
\, \psi_j \, dx
}
=
\frac{
\int_{-\infty}^{+\infty} \psi_2 \psi_3 \cdots \psi_{j-1}\psi_j
\, \hat{O}^{\dagger}\,\psi_1 \, dx
}{
\int_{-\infty}^{+\infty}
\psi_1 \psi_2 \psi_3 \cdots \psi_{j-1}
\, \psi_j \, dx
}
\label{eq:15}
\end{equation}
\end{widetext}
which reduces to the standard expectation value in the 3G limit. This formulation preserves the probabilistic interpretation while incorporating the modified phase structure induced by the NG geometry. The expectation value written in coordinate form as an integral is the position-space representation of the abstract bra–ket formulation of quantum mechanics. In the standard Hilbert-space picture, a quantum state is represented by a ket $|\psi\rangle$, and its dual by the bra $\langle\psi|$, with the inner product defined through spatial integration. The expression $\langle\psi|\hat O|\psi\rangle$ therefore corresponds, in the position basis, to the integral $\int \psi_{1}(x)\,\hat O\,\psi_{2}(x)\,dx$. Thus, the integral form of the expectation value is not a separate construction, but simply the coordinate realization of the underlying bra–ket structure. In the NG framework, the generalized expectation value retains this structural interpretation: it represents a modified pairing between a generalized bra and ket, expressed explicitly in coordinate space through the corresponding integral form. In this way, the bra-ket form of Eq.~\ref{eq:15} is given below:
\begin{equation}
\langle \hat{O} \rangle_{NG}
=
\frac{
\langle \psi_1 , \psi_2 , \ldots , \psi_{j-1} |
\, \hat{O}\,
| \psi_j \rangle
}{
\langle \psi_1 , \psi_2 , \ldots , \psi_{j-1}
| \psi_j \rangle
}
=
\frac{
\langle \hat{O} \psi_1
| \psi_2 , \ldots , \psi_j \rangle
}{
\langle \psi_1
| \psi_2 , \ldots , \psi_j \rangle
}
\label{eq:16}
\end{equation}
It is to be noted that in Eq.~\ref{eq:16}, the operator  $(\hat{O})$ only operates on one of the wavefunctions, either bra or ket form, as also shown in Eq. ~\ref{eq:15}. In the 2G framework, following this scheme, there is only the root of negative unity; therefore, the wave function does not have a complex conjugate. As an example, for the 4G framework, the probabilities of finding a particle are determined by multiplying the three wavefunctions, tnamely, the function $(\psi_{1} (x,t))$ itself and two complex conjugates, that is, $\psi_{2} (x,t)$ and $\psi_{3} (x,t)$. In the same way, for the 5G framework, the wavefunction $\psi_{5G} (x, t)$ of the particle has three complex conjugate in accordance with the roots of $(-1)^{1/4}$, namely $\Bar{\eta_{1}}$, $\Bar{\eta_{2}}$, $\Bar{\eta_{3}}$, and $\Bar{\eta_{4}}$. The resulting relation for the probability can then be written as $P_{5G}=\int_{-\infty}^{+\infty} \psi_{1} (x, t) \psi_{2} (x, t) \psi_{3} (x, t) \psi_{4} (x, t) dx$. The expectation values of an operator $\hat{O}$, in different dimensional geometries, can be expressed as $\langle\hat{O}\rangle_{2G}=\int_{-\infty}^{+\infty} \hat{O} \psi_{2G} (x, t) dx$, $\langle\hat{O}\rangle_{3G}=\int_{-\infty}^{+\infty} \psi_{1}(x, t) \hat{O} \psi_{2} (x, t) dx$, $\langle\hat{O}\rangle_{4G}=\int_{-\infty}^{+\infty} \psi_{1} (x,t) \psi_{2} (x,t) \hat{O} \psi_{3} (x, t) dx$, and $\langle\hat{O}\rangle_{5G}=\int_{-\infty}^{+\infty} \psi_{1} (x, t) \psi_{2} (x, t) \psi_{3} (x, t) \hat{O} \psi_{4} (x, t) dx$ for 2G, 3G, 4G, and 5G, respectively. The expressions for $\langle\hat{O}\rangle_{NG}$ can be used to prove that the Heisenberg principle $\Delta x_{NG} \Delta p_{NG} \geq \frac{\hbar}{2}$ holds for these geometries. This $L-^{j}$ integrable space-based scheme enables determining the Heisenberg uncertainty principle that must hold across a geometric framework of any dimension. By following the presented formulation, a heuristic form for calculating the average variance or uncertainty in the value of a hermitian operator $\hat{O}$ from normalized wavefunctions is given below: 
\begin{equation}
\begin{cases}
\Delta O = +0, & j=1 
\\[8pt]
\Delta O = \sqrt[2]{\langle\hat{O^{2}}\rangle - \langle\hat{O}\rangle^{2}}, & j=2
\\[8pt]
\Delta O \approx \sqrt[j]{\langle\hat{O^{j}}\rangle - \langle\hat{O}\rangle^{j}}
& j\ge 3 \\[8pt]
\end{cases}
\label{eq:17}
\end{equation}

In the NG framework for $j\ge 3$, the exact form of the uncertainty in $\hat{O}$ is governed by an $L^j$-norm structure, which can be found by starting with the following binomial expansion:
\begin{equation}
(O-\langle O\rangle)^j = \sum_{k=0}^{j} \binom{j}{k} (-1)^k O^{j-k}\langle O\rangle^k,
\end{equation}
and taking expectation values for the $j$-th central moment,
\begin{equation}
\left\langle (O-\langle O\rangle)^j \right\rangle
=
\sum_{k=0}^{j} \binom{j}{k} (-1)^k \langle O^{j-k}\rangle \langle O\rangle^k.
\end{equation}
For specific cases of $j=3$, and $j=4$, this yields $\left\langle (O-\langle O\rangle)^3 \right\rangle
=
\langle O^3\rangle -3\langle O^2\rangle\langle O\rangle +2\langle O\rangle^3$, and $\left\langle (O-\langle O\rangle)^4 \right\rangle
=
\langle O^4\rangle -4\langle O^3\rangle\langle O\rangle +6\langle O^2\rangle\langle O\rangle^2 -3\langle O\rangle^4$, respectively. These expressions reduce to $\Delta O \approx \sqrt[3]{\langle\hat{O^{3}}\rangle - \langle\hat{O}\rangle^{3}}$, and $\Delta O \approx \sqrt[4]{\langle\hat{O^{4}}\rangle - \langle\hat{O}\rangle^{4}}$ provided $\langle O^2\rangle=\langle O\rangle^{2}$ for $j=3$, and $\langle O^3\rangle=\langle O\rangle^{3}$ and $\langle O^2\rangle=\langle O\rangle^{2}$ for $j=4$, respectively. The generalized variance herein involves not only the difference of moments but also mixed terms that encode correlations between different orders. However, within the NG framework, where the probability measure and expectation values are defined through a $j$-fold product structure, the leading contribution to statistical dispersion is naturally captured by the difference of moments $\left\langle (O-\langle O\rangle)^j \right\rangle \approx \langle O^j\rangle - \langle O\rangle^j$, so as to define the generalized uncertainty (or $j$-th order variance) as $\Delta O_{NG} = \left(\langle O^j\rangle - \langle O\rangle^j\right)^{1/j}$. In this sense, the generalized variance represents the average spread of the observable in NG geometries, preserving dimensional consistency while reflecting the higher-order geometric structure of the underlying space. Additionally, it is critical to verify the validity of Heisenberg's uncertainty principle for NG frameworks, which is done in the latter sections by determining the usual uncertainties in momentum and position, i.e., $\Delta x \Delta p \geq \frac{\hbar}{2}$. As mentioned earlier, it will be carried out by determining the uncertainties in quantum-mechanical operators in the $L^{j}$- integrable space.

In the context of the NG frameworks, the canonical commutation relations, such as between the position and momentum of a quantum particle in a one-dimensional potential, are also important. These relations, along with their corresponding SEs, are presented in the latter sections for the 2G, 3G, 4G, and 5G frameworks. It has been shown that the canonical commutator relations for position and momentum can be generalized to the NG framework in the following way: 

\begin{equation}
\left[x,p\right]_{NG}= (-1)^{1/j} \hbar
\end{equation}  

The generalized commutation relation extends the canonical structure of quantum mechanics to $N$-dimensional geometric (NG) frameworks. Unlike the standard 3G case, where $[x,p] = i\hbar$, the commutator acquires a geometry-dependent phase factor given by the $j$-th root of negative unity. This reflects the underlying multi-phase structure of the NG formulation, in which the algebra of observables is governed by a set of complex roots determined by the geometry. Despite this modification, the commutation relation preserves the fundamental role of $\hbar$ as the scale of quantum fluctuations, ensuring consistency with generalized uncertainty relations across different geometric dimensions.

\subsection{Quantum Free States in Multidimensional Geometric Frameworks}
A fundamental test of the generalized NG Schr\"odinger equation is the case of a free quantum particle, for which the potential energy vanishes, $V_{NG}(x) = 0$. In this situation, the dynamics are entirely governed by the generalized momentum and energy operators derived from the modified dispersion relation in Eq.~\ref{eq:7}. The time-dependent NG Schr\"odinger equation for a free particle in the NG framework can therefore be written as
\begin{equation}
-\frac{1}{j}\frac{\hat{p}^{\,j}}{m^{\,j-1}c^{\,j-2}} \,
\psi_{NG}(x,t)
=
\hat{E}\,\psi_{NG}(x,t),
\end{equation}
where $j = N-1$ and $\hat{p}^{\,j}$ denotes the $j$-th order momentum operator consistent with the NG framework. Seeking stationary solutions of the separable form of $\psi_{NG}(x,t)
=
\phi_{NG}(x)\,\Theta(t)$ leads to the spatial eigenvalue equation $\hat{p}_{NG}\,
\phi_{NG}(x)
=
p_{NG}\phi_{NG}(x),$ and $\hat{E}_{NG} \Theta_{NG}(t)
=
E_{NG} \Theta_{NG}(t)$ for $\hat{p}$, and $\hat{E}$ hermitian operators, respectively. As the momentum operator remains proportional to spatial derivatives, plane-wave solutions persist in the NG framework as an oscillatory function whose argument has to be of the form
$\left(a_{j}~{k}{x} \right)$ in which $a_{j}$ are the roots of negative unity. Thus, resulting in the following form of the spatial form of the wave function: 
\begin{equation}
\phi_{NG}({x})
=
C \exp\!\left( j^{th}~root~of~(-1)^{\frac{1}{j}}\,{k_{NG}}{x} \right),
\label{eq:22}
\end{equation}
and the corresponding temporal part as:
\begin{equation}
\Theta(t)_{NG} =
\begin{cases}
A \exp\!\left(
\,\frac{j^{th}~root~of~\,(-1)^{1/j}\,E_{NG}}{\hbar}\,t \right), & j = 1, 
\\[9pt]

A \exp\!\left(
-\,\frac{j^{th}~root~of~\,(-1)^{1/j}\,E_{NG}}{\hbar}\,t
\right), & j \ge 2.
\end{cases}
\end{equation}
\label{theta}

The application of the $\hat{p_{NG}}$, and $\hat{E_{NG}}$ operators onto their corresponding wave functions of Eq.\ref{eq:22}, and Eq.~\ref{theta} yields the eigenvalues of these operators, which are  given below as:
\begin{equation}
\begin{cases}
p_{NG} = \hbar k_{NG}, 

\\[8pt]

E_{NG} = \frac{1}{j}\frac{\hbar^{j} k_{NG}^{j}}{m^{j-1}c^{j-2}}
\end{cases}
\label{eq:24}
\end{equation}

and these reproduce the familiar quadratic relation $E_{NG} = \hbar^{2}{k_{NG}}^{2}/2m$ recovered in the 3G case ($j=2$). Thus, while the functional form of the plane-wave solution remains structurally similar to that of standard quantum mechanics, the energy now scales as a higher power of the wave number. This result demonstrates that translational invariance is preserved within the NG framework, and that free-particle states continue to be characterized by well-defined momentum eigenvalues. However, the modified dispersion relation implies that the group velocity, phase velocity, and energy–momentum scaling differ fundamentally from those in the 3G case. In particular, the energy grows as ${k_{NG}}^{\,j}$ rather than quadratically, leading to distinct propagation characteristics for $j \neq 2$.

Hence, the quantum free state $\psi_{NG}(x,t)$ of the particle traveling to the positive axis in the 2G framework can be expressed as follows:
\begin{equation}
\psi_{2G}(x, t)= \mathbf{N_{2G}} \left(e^{-k_{2G}x} \right) \left[\exp{\left(-\frac{E_{2G}}{\hbar} t\right)}\right]
\label{eq:25}
\end{equation}

With:
\[
\mathbf{N_{2G}} = k_{2G}, \quad \text{for } 0 \le x < \infty.
\]

Whereas for $j\ge2$, the quantum free state becomes
\begin{widetext}
\begin{equation}
\psi_{NG}(x, t)= \mathbf{N_{NG}} \exp\!\left( j^{th}~root~of~(-1)^{\frac{1}{j}}\,{k_{NG}}{x} \right) \exp\!\left(
-\,\frac{j^{th}~root~of~\,(-1)^{1/j}\,E_{NG}}{\hbar}\,t
\right),
\label{eq:26}
\end{equation}  
\end{widetext}
With:
\[
\mathbf{N_{NG}} = \text{(j-1) roots of}~(-1)^{1/j}~ k_{NG}, \quad \text{for } 0 \le x < \infty.
\]
Hence, translational invariance and the plane-wave structure are preserved, but the energy–wave-number scaling and the associated phase and group velocities become geometry-dependent. These free-state solutions provide the foundation for the bound-state analysis in the next section, where confinement and boundary conditions lead to quantized spectra and geometry-specific eigenfunctions.

\subsection{Quantum Bound States in Multidimensional Geometric Frameworks}
To examine the consequences of the generalized Schr\"odinger equation derived for the NG framework, we now consider a quantum particle confined within a one-dimensional infinite potential well. This system provides the simplest nontrivial setting in which quantization arises purely from boundary conditions, independent of the detailed functional form of interaction potentials \cite{ReedSimon1975, Schmudgen2012, vonNeumann1955}. In this case, the potential energy is defined as $V_{NG}(x)=0$ within a finite region of length $l$ and $V_{NG}(x)=\infty$ outside that region, thereby restricting the particle to a bounded spatial domain. In the case of the infinite potential well, the confinement of the particle is implemented through Dirichlet boundary conditions imposed on the wavefunction at the edges of the domain. Specifically, for a particle confined to the interval $0 < x < l$, the boundary conditions (BCs) $\psi(0,t) = 0, \qquad \psi(l,t) = 0$ are enforced, i.e., combined Dirichlet boundary conditions \cite{Munschy2024, Guedes2025}. These BCs follow from the requirement that the wavefunction vanish in regions where the potential is infinite, ensuring that the particle has zero probability of being found outside the well. In the standard 3G formulation, Dirichlet boundary conditions guarantee that the Hamiltonian operator with domain restricted to functions vanishing at the endpoints is self-adjoint on $L^2(0,l)$, leading to a real and discrete energy spectrum \cite{Shankar1994,SakuraiNapolitano2017}. In the present NG framework, the same boundary conditions are adopted as the natural generalization of confinement within a finite spatial domain, allowing direct comparison with the conventional quadratic case. The infinite potential well serves as a model-independent framework for analyzing how the modified kinetic operator associated with $j = N-1$ alters the structure of eigenfunctions, eigenenergies, and spectral scaling relative to the conventional 3G case. By imposing appropriate boundary conditions on the generalized differential equations, the quantization emerges in different NG geometries, namely wave vector $k_{NG}$ in the form of quantized $k_{NG,n}$, and the energy of particles $E_{NG}$ in the quantized energy  $E_{NG,n}$.The corresponding quantized or bound states of the particles, such as in the case of the particles in a potential well, the $\phi_{NG,n}(x)$ can be determined from the generalized SE in the NG framework. We present next a scheme for determining these bound states of a particle in an infinite potential well: 
\begin{itemize}
\item Find the $j=N-1$ roots of the negative unity for a given NG framework. For instance, for the case of the 2G framework, the $j$ roots of the negative unity are only one because $j=1$, which is $a_{1}=-1$, or $a_{1}=\Bar{1}$. Whereas the $j$ square roots of the negative unity are $a_{1}=\iota$ and $a_{2}=-\iota$ for the 3G case, since $j=2$. The negative roots of unity ($a=(-1)^\frac{1}{j}$) for even and odd N-dimensional geometric frameworks can be written as $a_{j}=e^{\pi \iota \frac{2q+1}{j}}$, and $a_{j}=e^{2\pi \iota \frac{q}{j}}$, respectively with, $q=0, 1, 2, ...j-1$. 
    \item Express $f(kx)$ for even and odd N-dimensional geometries as a linear combination of exponential functions by multiplying their arguments with the negative of the roots of negative unity. Mathematically, this can be written as follows:
for the N-dimensional case:
    \[
    f(kx) =  {\sum_{j=1}^{N-1}} C_{j} \exp(a_{j} kx)
    \]
      
    For instance, this relation correspondingly reduces in the following for 2G and 3G frameworks. 
    \[
    f(kx) = \left[ C_{1} \exp(\Bar{1} kx) \right]
    \]
    and
    \[
    f(kx) = \left[ C_{1} \left( \exp(-\iota kx) \right) + C_{2} \left( \exp(\iota kx) \right) \right]
    \]
    for 2G and 3G, respectively. The $C_{1}$, $C_{2}$, and $C_{3}$ are arbitrary constant coefficients.

    \item Plug the value of $f(kx)$ into the equation $\phi_{n}(x)= f(kx)$. Next, simplify the exponential terms using Euler's identity. 
    \item Apply combined Dirichlet BCs, i.e., $f(x=0)=0=f(x=l)$ for a potential well with infinite voltage height and width of $l$ in length, to $f(kx)$ to determine the coefficients and conditions in $k_{NG}$ that turn it into $k_{NG,n}$ for the stationary part of the wavefunctions $(\phi_{NG,n})$ with $n=1, 2, 3, ....$.
    \item Determine the time-dependent part $(\Theta_{NG} (t))$ of the wavefunctions using the generalized operator form for the energy of particles.  
    \item Normalize the wavefunctions ($\phi_{NG, n} (kx))$ and calculate the expectation values ($\langle\hat{O}\rangle_{NG}$) of the quantum mechanical quantities. This can be done by assuming that the wavefunctions are $j$ integrable for the NG framework, which means $\int \phi_{NG,n}^{j} dx \leq \infty$. This is further generalized in the "discussion" section. 
\end{itemize}


Special cases of the NG framework for a quantum particle in a potential well are explored next. The forms of the SE in 2G, 3G, 4G, and 5G are determined for a particle in a one-dimensional potential well with $V_{NG}=\infty$ for $x<0$ and $x>l$, and $V_{NG}=0$ for $0<x<l$. The states or wavefunctions obeying the SE in the $\mathbb{H}^n$ Hilbert vector space are determined for these geometries.

\begin{figure}[t]
    \centering
    \includegraphics[width=0.42\linewidth]{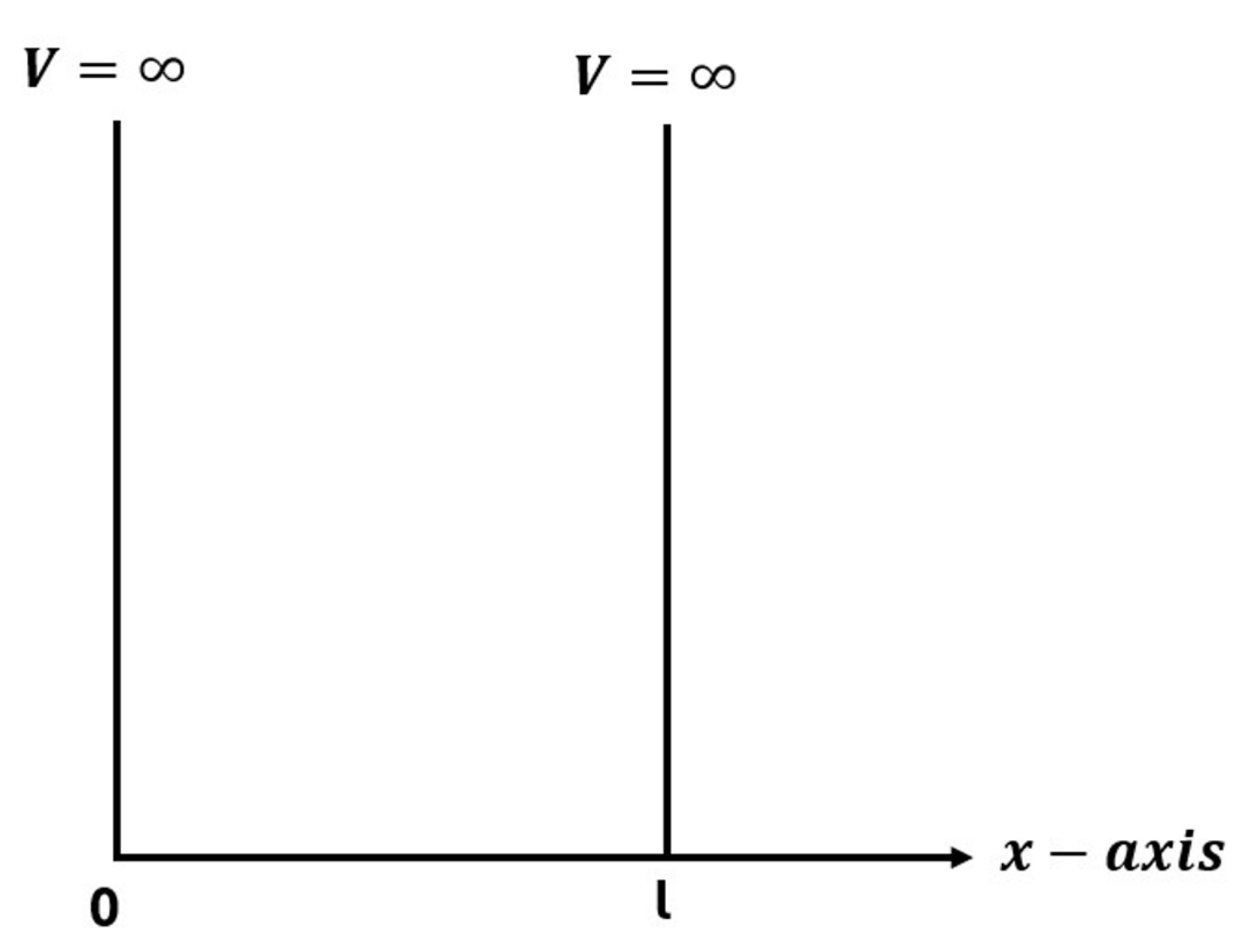}
    \caption{Schematic representation of a quantum particle confined in a one-dimensional infinite potential well of width $l$, with $V_{NG}=0$ for $0<x<l$ and $V_{NG}=\infty$ outside.}
    \label{fig:potential_well}
\end{figure}

\subsubsection{Particle in a One-Dimensional Potential Well in the 2G Framework}
For the 2G framework $(j=1)$, the Eq. ~\ref{eq:10} for a free particle moving only along the x-axis's SE can be written in the following form:
\begin{equation}
\hbar c \frac{\partial}{\partial x} \psi_{2G}= \hbar \frac{\partial}{\partial t} \psi_{2G}(x,t) 
\end{equation} 

Let us denote $\psi_{2G} (x, t)$ as $\psi (x, t)$ for short. Furthermore, it $\psi(x.t)$ can be expressed as $\psi(x,t) = \phi_{n}(x) \Theta(t)$ into separate spatial and temporal parts. In this way, the time-independent form of the SE can be written in the following way: 
\begin{equation}
-\hbar c \frac{\partial}{\partial x} \phi_{2G,n}(x) = E_{2G,n} \phi_{2G,n}(x) 
\end{equation} 

The temporal part $(\Theta_{2G} (t))$ of $\psi_{2G}$ varies in time as "$\ \exp\left({-\frac{E_{2G}}{\hbar}t}\right)$". The solution for the spatial part can be taken as an exponential one, with the argument multiplied by the negative linear root of negative one. It means that $\phi_{2G,n}(x)$ can be written as $f(kx)$ in the form below. 

\begin{equation}
f(kx)=C_{1}\exp{\left(-kx\right)}
\end{equation}

There is only one term in the above equation, and its Euler form does not exist because its argument does not contain a complex number ($\iota$). The eigenfunctions of a particle confined in a potential well below limits such as $V_{2G}=\infty$ for $x<0 \ and \ x>l$ and $V_{2G}=0$ for $0<x<l$, can be found by applying the boundary conditions (BC) in $\phi_{2G,n}(x)= f(kx)$, as $\phi_{2G,n}(x=0)=0=\phi_{2G,n}(x=l)$. After doing this, it was found that it yields a trivial solution, i.e., $\phi_{2G} (x)=0$. This implies that, in the 2G framework, the particle in a potential well does not have bound states. As stated earlier, the wavefunction $\psi_{2G}(x,t)$ represents the state of an unbounded particle traveling to the positive axis in the 2G framework, which can be written in the following way:
\begin{equation}
\psi_{2G}(x, t)= \mathbf{N_{2G}} \left(e^{-k_{2G}x} \right) \left[\exp{\left(-\frac{E_{2G}}{\hbar} t\right)}\right]
\label{eq:30}
\end{equation}

With;
\begin{equation}
\mathbf{N_{2G}}= \begin{cases}
k_{2G}, & \text{for } 0 \le x \le \infty, \\[8pt]
\dfrac{k_{2G}}{1 - e^{-k_{2G}l}}, & \text{for } 0 \le x \le l.
\end{cases}
\end{equation}

And the eigen energy $E_{2G}$:
\begin{equation}
E_{2G} = \hbar k_{2G}c 
\end{equation} 

In the 2G framework, the wavefunction given in Eq.~\ref{eq:30} resembles the quantum free particle in Eq.~\ref{eq:25}, which also exhibits exponential decay in both space and time rather than oscillatory behavior. Unlike the conventional 3G case, where the presence of imaginary phases leads to bounded sinusoidal solutions, the 2G formulation involves only real exponential functions arising from the linear root of negative unity. As a result, the spatial component \(\exp(-k_{2G}x)\) decays monotonically with \(x\), while the temporal factor \(\exp(-E_{2G}t/\hbar)\) also decreases with time. This behavior reflects the underlying linear dispersion relation \(E_{2G} = \hbar k_{2G} c\), which does not support standing-wave solutions under Dirichlet boundary conditions. Consequently, no discrete eigenvalues \(k_{2G,n}\) emerge, and the system does not admit normalizable bound states. The non-oscillatory and decaying nature of the wavefunction is therefore consistent with the interpretation that, in the 2G framework, the particle behaves as a freely propagating, massless-like excitation rather than forming quantum bound states. Nevertheless, the wavefunction of the quantum particle in the 2G framework does obey the relation of the Poisson bracket for $j=1$ that works for $\hat{x}$, and $\hat{p}$ in the following way. 

\begin{equation}
\left[x,p\right]_{2G}= (-1)^{1/1} \hbar
\end{equation}

\subsubsection{Particle in one-dimensional Potential Well in the 3G Framework}
Although it is redundant to include a quantum-mechanical formulation in 3G \cite{GriffithsSchroeter2018}, it is imperative to do so to establish connections with quantum mechanics for other cases, including 2G, 4G, and 5G. The time-dependent S.E. in the 3G framework can be written as given below: 

\begin{equation}
 -\frac{\hbar^{2}}{2m} \frac{\partial^{2}}{\partial x^{2}} \psi_{3G} (x,t) = \iota \hbar \frac{\partial}{\partial t} \psi_{3G} (x,t) 
\end{equation} 

The motion of a particle in a potential well in the 3G framework is the same as discussed in the conventional quantum theory of particles, described in the SE in the 3G framework. The eigenfunctions $\left(\psi_{3G}\right)$ and eigenenergy $\left(E_{3G,n}\right)$ of a particle in a one-dimensional potential box can be given below: 

\begin{equation}
\psi_{3G, n} (x, t) = f(kx) \left[\exp\left(-\iota {\frac{E_{3G,n}}{\hbar}t}\right)\right] 
\end{equation}

The form of $\phi_{3G,n}(x)= f(kx)$ needs to be determined that obeys the BCs, that is, $\phi_{3G,n} (x=o)=0=\phi_{3G,n} (x=l)$. It can be shown that the $f(kx)$ has two possible solutions, which can be generated by multiplying the arguments of exponential functions with the negative of the square roots of negative unity, namely, $\exp{\left(-\iota kx\right)}$, and $\exp{\left(-\Bar{\iota} kx\right)}$, with $\Bar{\iota}=-\iota$. In this way, a general solution for $f(kx)$ will be a linear combination of these eigenfunctions in the following form: 

\begin{equation}
 f(kx)= C_{1} exp{\left(-\iota kx\right)} + C_{2} \exp{\left(-\Bar\iota kx\right)}
\end{equation} 

By plugging in the relations of these cube roots of negative unity and subsequently using the Euler identity, it can be shown that $f(kx)$ has the following form: 

\begin{equation}
 f(kx)= 2 \left(C_{1}+C_{2}\right) cos\left(kx\right) + 2 \iota\left(C_{1}-C_{2}\right) sin\left(kx\right)
\end{equation}

The application of BCs, i.e., $\phi_{3G, n}(x=0)=0$, and $\phi_{3G, n}(x=l)=0$, allows determining the coefficients $C_{1}$, and $C_{2}$ and it can be found out that $C_{1}=-C_{2}$, and $k_{3G, n} l= n \pi$.  This will result in the following form for $\phi_{3G,n}(x)$:

\begin{equation}
 \phi_{3G, n} (x) = \mathbf{N_{3G, n}} sin \left(\frac{n \pi}{l} x\right) 
\end{equation}

The constants $2 \iota(C_{1}-C_{2})$ have been equated to another arbitrary constant $\mathbf{N_{3G, n}}$, which can be determined using the normalization condition. In this way, the eigenfunction and eigenenergy of the particle in a potential well in the 3G framework can be written in the following way.  

\begin{equation}
\psi_{3G, n} (x, t) = \sqrt{\frac{2}{l}} \left[sin \left(\frac{n \pi}{l} x\right)\right] \left[\exp{\left(-\iota \frac{E_{3G, n}}{\hbar}t\right)}\right] 
\end{equation}

And the eigenenergy $E_{3G, n}$:
\begin{equation}
E_{3G, n} = \frac{\hbar^{2} k^{2}_{3G, n}}{2m}=n^{2} \frac{\hbar^{2} \pi^{2}}{2ml^{2}} 
\end{equation} 

In the same way, the Poison bracket that the wavefunctions of the particle in 3G geometry obey can be written as: 
\begin{equation}
\left[x,p\right]_{3G} = (-1)^{1/2} \hbar \equiv \iota \hbar
\end{equation} 

\subsubsection{Particle in one-dimensional Potential Well in the 4G Framework}

The 4G framework, corresponding to \( j = 3 \), represents the first nontrivial extension beyond the conventional quadratic (3G) structure, introducing a cubic kinetic operator into the Schr\"odinger equation. This modification fundamentally alters both the differential order of the governing equation and the resulting spectral properties of confined quantum systems. In contrast to the 3G case, where oscillatory solutions arise purely from complex conjugate pairs, the 4G framework incorporates the cube roots of negative unity, leading to a richer structure that combines oscillatory and exponentially modulated behavior. As a result, the eigenfunctions acquire mixed exponential--trigonometric forms, while the energy spectrum exhibits cubic scaling with the quantum number. This section systematically derives the wavefunctions and eigenenergies for a particle confined in a one-dimensional infinite potential well within the 4G geometry, highlighting how the generalized kinetic operator reshapes quantization and boundary-condition constraints.
\begin{equation}
 -\frac{\hbar^{3}}{3m^{2}c} \frac{\partial^{3}}{\partial x^{3}} \psi_{4G} (x,t) = \bar{\omega}^{2} \hbar \frac{\partial}{\partial t} \psi_{4G} (x,t) 
\end{equation} 
The motion of a particle in a potential well is an interesting case for understanding quantum mechanics in the 4G framework. The quantized wavefunctions $\left(\psi_{4G, n}\right)$ and energy $\left(E_{4G, n}\right)$ of a particle in a one-dimensional potential box can be given below: 
\begin{equation}
\psi_{4G, n} (x, t) = \phi_{4G, n}(x) \Theta (t)= f(kx) \left[\exp{\left(-\bar{\omega}^{2} {\frac{E_{4G, n}}{\hbar}t}\right)}\right] 
\end{equation}

The $k$ is the particle wave vector in the 4G framework. In the 4G framework, the probability of finding the particle in the potential well should be a real number, as verified by Eq.~\ref{eq:14}. For a particle confined in a potential well within limits like $V_{4G}=\infty$ for $x<0 \ and \ x>l$, and $V_{4G}=0$ for $0<x<l$, the following boundary conditions can be applied to the spatial part of the eigenfunctions,$\phi_{4G, n}(x=0)=0=\phi_{4G, n}(x=l)$ that obey the following time-independent SE. 
\begin{equation}
- \frac{\hbar^{3}}{3m^{2}c} \frac{\partial^{3}}{\partial x^{3}} \phi_{4G, n} (x) = E_{4G, n} \phi_{4G, n} (x) 
\end{equation} 
This implies that the form of $\phi_{4G, n}(x)= f(kx)$ needs to be determined to obey the boundary conditions listed above (BCs). It can be shown that $f(kx)$ has three possible solutions, which can be generated by multiplying the arguments of exponential functions with negatives of the cube root of negative unity, namely $\exp{\left(-\Bar{1} kx\right)}$, $\exp{\left(-\Bar{\omega_{1}} kx\right)}$, and $\exp{\left(-\Bar{\omega_{2}} kx\right)}$. In this way, a general solution $f(kx)$ will be a linear combination of these eigenfunctions in the following form: 
\begin{widetext}
\begin{equation}
 f(kx)= C_{1} exp{\left(\Bar{1} kx\right)} + C_{2} exp{\left(\Bar{\omega_{1}} kx\right)} + C_{3} \exp{\left( \Bar{\omega_{2}} kx\right)}
\end{equation} 
\end{widetext}
By plugging in the relations of these cube roots of negative unity and subsequently using the Euler identity, it can be shown that $f(kx)$ has the following form: 
\begin{widetext}
\begin{equation}
 f(kx)= C_{1} exp{\left(-kx\right)} + \left(C_{2}+C_{3}\right) exp{\left(kx/2\right)} cos\left(\frac{\sqrt{3}}{2} kx\right) + \iota \left(C_{2}-C_{3}\right) exp{\left(kx/2\right)} sin\left(\frac{\sqrt{3}}{2} kx\right)
\label{4G}
\end{equation}
\end{widetext}
The application of the BC of $\phi_{4G, n}(x=0)=0$, allows determining a relation among $C_{1}$, $C_{2}$, and $C_{3}$ coefficients, which turn turned out to be $C_{1}=-(C_{2}+C_{3})$. Even though the application of $\phi_{4G, n}(x=l)=0$ allows setting $\frac{\sqrt{3}}{2} k_{4G, n} l= n \pi$ for the third term in Eq.~\ref{4G}, i.e.,$\iota \left(C_{2}-C_{3}\right) exp{\left(kx/2\right)} sin\left(\frac{\sqrt{3}}{2} kx\right)$, but first two terms, i.e., $C_{1} exp{\left(-kx\right)} + \left(C_{2}+C_{3}\right) exp{\left(kx/2\right)} cos\left(\frac{\sqrt{3}}{2} kx\right)$ do not reduce to $0$ for $\frac{\sqrt{3}}{2} k_{4G, n} l= n \pi$. This means that, in the 4G framework, the most general solution given in Eq.~\ref{4G} does not lead to a bound state for a particle in a box. However, the solution given in Eq.~\ref{4G} without the first term does lead to the bound state for a particle in a box in the 4G framework. It can be shown first by writing Eq.~\ref{4G} without the term takes as given below: 
\begin{widetext}
\begin{equation}
 f(kx)= \left(C_{2}+C_{3}\right) exp{\left(kx/2\right)} cos\left(\frac{\sqrt{3}}{2} kx\right) + \iota \left(C_{2}-C_{3}\right) exp{\left(kx/2\right)} sin\left(\frac{\sqrt{3}}{2} kx\right)
\label{4G-2}
\end{equation}
\end{widetext}
, and the application of the Dirichlet BCs leads to the following form for $\phi_{4G, n}(x)$:

\begin{equation}
 \phi_{4G, n} (x) = \mathbf{N_{4G, n}} exp{\left(n\frac{\pi}{ \sqrt{3}l} x\right)} sin\left(n\frac{\pi}{l} x\right) 
\end{equation}

The constant \(2\iota C_2\) has been equated to another arbitrary constant $\mathbf{N_{4G, n}}$, which can be determined using the normalization condition. This normalization ensures that the generalized probability density, defined within the \(L^{j=3}\)-norm framework, remains finite and physically meaningful over the system's domain. In particular, the value of $\mathbf{N_{4G, n}}$ is fixed by requiring that the integrated probability over the interval \(0 < x < l\) equals unity, consistent with the generalized probabilistic interpretation introduced earlier. In the 4G framework, we can write the particle’s eigenfunction and eigenenergy as follows:
\begin{widetext}
\begin{equation}
\psi_{4G, n} (x, t) = \mathbf{N_{4G, n}} \left[exp{\left(n\frac{\pi}{ \sqrt{3}l} x\right)} sin\left(n\frac{\pi}{l} x\right)\right] \left[\exp{\left(-\bar{\omega}^{2} {\frac{E_{4G, n}}{\hbar}t}\right)}\right] 
\end{equation}
\label{eq:49}
\end{widetext}
With
\begin{equation}
\mathbf{N_{4G, n}} = \left[ \frac{8 n \pi}{\,l\left(1 - (-1)^n e^{\sqrt{3}\, n \pi}\right)} \right]^{1/3}
\label{eq:50}
\end{equation}

and; 
\begin{equation}
E_{4G, n} = \frac{\hbar^{3} k^{3}_{4G, n}}{3m^{2}c}=n^{3} \left(\frac{8\hbar^{3} \pi^{3}}{27 \sqrt{3} m^{2}cl^3}\right) 
\label{eq:51}
\end{equation}

The vanishing of the wavefunction at the boundary \(x = l\) arises from the oscillatory sine function rather than the exponential term. While the exponential component introduces a spatial damping that suppresses the amplitude across the domain, it does not enforce the boundary condition itself. Instead, the node at \(x = l\) is determined by the cosine term, indicating that quantization is governed by the oscillatory structure, whereas the geometric modification introduces an asymmetric, exponentially attenuated envelope. In the same way, the Poison bracket that the wavefunctions, $\phi_{4G, n} (x)$ or $\psi_{4G, n} (x)$, obey can be written as: 
\begin{equation}
\left[x,p\right]_{4G} = (-1)^{1/3} \hbar \equiv \Bar{\omega_{1}} \hbar
\end{equation} 

\subsubsection{Particle in one-dimensional Potential Well in the 5G Framework}
The 5G framework, corresponding to \( j = 4 \), extends the generalized formulation to a quartic kinetic structure, further amplifying the departure from the conventional 3G quantum mechanics. In this case, the Schr\"odinger equation involves fourth-order spatial derivatives, reflecting the underlying quartic dispersion relation. The presence of four roots of negative unity introduces an even richer phase structure, resulting in eigenfunctions that exhibit a combination of hyperbolic and trigonometric behavior. This leads to spatial profiles characterized by both oscillatory components and strong exponential modulation. Consequently, the quantization conditions and admissible solutions become more intricate, while the energy spectrum scales as the fourth power of the quantum number, \( n^4 \). This section develops the explicit forms of the wavefunctions and eigenenergies for a particle confined to a one-dimensional infinite potential well within the 5G framework, illustrating how higher-order geometry systematically modifies both spectral growth and the functional structure of quantum states.

\begin{equation}
 -\frac{\hbar^{4}}{4m^{3}c^{2}} \frac{\partial^{4}}{\partial x^{4}} \psi_{5G} (x,t) = \bar{\eta_{4}} \hbar \frac{\partial}{\partial t} \psi_{5G} (x,t) 
\end{equation} 

The quantum confinement of a particle in a potential well turns out ot be an interesting case for understanding quantum mechanics in the 5G framework. The eigenfunctions $\left(\psi_{5G, n}\right)$ and eigenenergy $\left(E_{5G, n}\right)$ of a particle in a one-dimensional potential box can be given below: 
\begin{equation}
\psi_{5G,n}(x,t)=\phi_{5G,n}(x)\,\Theta(t)
= f(kx)\exp\!\left(-\bar{\eta}_{4}\,\frac{E_{5G,n}}{\hbar}\,t\right)
\end{equation}

The $k$ is the particle wave vector in the 5G framework. The probability $P$ of finding the particle in the potential well should be a real number, determined according to the methods presented in the later sections. For a particle confined in a potential well within limits like $V=\infty$ for $x<0 \, \text{and} \, x>l$, and $V=0$ for $0<x<l$, the following boundary conditions can be applied to the spatial part of the eigenfunctions,$\phi_{5G, n}(x=0)=0=\phi_{5G, n}(x=l)$ that obey the following time-independent SE. 
\begin{equation}
 -\frac{\hbar^{4}}{4m^{3}c^{2}} \frac{\partial^{4}}{\partial x^{4}} \phi_{5G, n} (x) = E_{5G, n} \phi_{5G, n} (x) 
\end{equation} 
This implies that the form of $\phi_{5G, n}(x)= f(kx)$ needs to be determined to obey the boundary conditions listed above (BCs). It can be shown that $f(kx)$ has three possible solutions, which can be generated by multiplying the arguments of exponential functions with the negative of the quartic roots of negative unity, namely, $\exp{\left(\Bar{\eta_{1}} kx\right)}$, $\exp{\left(\Bar{\eta_{2}} kx\right)}$, $\exp{\left(\Bar{\eta_{3}} kx\right)}$, and $\exp{\left(\Bar{\eta_{4}} kx\right)}$. In this way, a general solution for $f(kx)$ will be a linear combination of these eigenfunctions in the following form: 

\begin{widetext}
\begin{equation}
 f(kx)= C_{1} \exp{\left(\Bar{\eta_{1}} kx\right)} + C_{2} \exp{\left(\Bar{\eta_{2}} kx\right)} + C_{3} \exp{\left(\Bar{\eta_{3}} kx\right)} + C_{4} \exp{\left(\Bar{\eta_{4}} kx\right)}
 \label{5G}
\end{equation} 
\end{widetext}

After plugging in the relations of roots of the negative unity quartet and subsequently using the Euler identity, $f(kx)$ takes the following form: 
\begin{widetext}
\begin{equation}
\begin{split}
f(kx) = & \left[ \left(C_{1} + C_{2}\right) \exp\left(\frac{kx}{\sqrt{2}}\right) + \left(C_{3} + C_{4}\right) \exp\left(-\frac{kx}{\sqrt{2}}\right) \right] \cos\left(\frac{kx}{\sqrt{2}}\right) \\
& + \iota \left[ \left(C_{1} - C_{2}\right) \exp\left(\frac{kx}{\sqrt{2}}\right) + \left(C_{3} - C_{4}\right) \exp\left(-\frac{kx}{\sqrt{2}}\right) \right] \sin\left(\frac{kx}{\sqrt{2}}\right)
\end{split}
\label{eq:57}
\end{equation}
\end{widetext}

The application of the BC of $\phi_{5G,n} (x=0)=0$, leads to the $C_{1}+C_{2}=-(C_{3}+C_{4})$. In addition, the waves with coefficients $C_{1}$ and $C_{3}$ ,and with coefficients $C_{2}$ and $C_{4}$ are traveling in opposite directions to each other. By assuming these are the waves with the same amplitudes, but can be either in phase or out of phase. The out-of-phase condition means that $C_{1}= -C_{3}$ and $C_{2}=-C_{4}$. Using these conditions, the Eq.~\ref{eq:57} can be written in the following form: 
\begin{widetext}
\begin{equation}
f(kx) = 2\left[\left(C_{1} + C_{2}\right) \sinh\left(\frac{kx}{\sqrt{2}}\right)\right] \cos\left(\frac{kx}{\sqrt{2}}\right)  + 2\iota \left[ \left(C_{1} - C_{2}\right) \sinh\left(\frac{kx}{\sqrt{2}}\right) \right] \sin\left(\frac{kx}{\sqrt{2}}\right)
\label{eq:58}
\end{equation}
\end{widetext}

The application of the second BC, i.e., $\phi_{5G, n}(kl)=0$, leads to the oscillating bound states for the particle in a box in the 5G framework with $\frac{1}{\sqrt{2}} k_{5G, n} l=n \pi$. This results in the following form for $\phi_{5G, n}(x)$:

\begin{equation}
 \phi_{5G, n} (x) = \mathbf{N_{5G, n}} \sinh\left(\frac{n\pi} {l} x\right) \sin\left(\frac{n\pi} {l} x\right) 
 \label{eq:59}
\end{equation}

The constants $2\iota (C_{1}-C_{2})$ have been equated to another arbitrary constant $\mathbf{N_{5G, n}}$, which can be determined using the normalization condition. It should be noted that $n=1, 2,3,... $ is an integer that describes the quantum states of a particle in the potential well. In this way, the eigenfunction and eigenenergy of the particle in a potential well in the 5G framework can be written the following way: 
\begin{widetext}
\begin{equation}
\psi_{5G, n} (x, t) = \mathbf{N_{5G, n}} \sinh\left(\frac{n\pi} {l} x\right)  \sin\left(\frac{n\pi} {l} x\right)  \left[\exp{\left(-\bar{\eta_{4}} {\frac{E_{5G, n}}{\hbar}t}\right)}\right] 
\label{eq:60}
\end{equation}
\end{widetext}

With,
\begin{widetext}
\begin{equation}
\mathbf{N_{5G, n}} = \left[\frac{2560\,n\pi}{l\left(360\pi n - 96\sinh(2\pi n) + 3\sinh(4\pi n)\right)}\right]^{1/4}
\label{eq:61}
\end{equation}
\end{widetext}

and; 
\begin{equation}
E_{5G, n} = \left(\frac{\hbar^{4} k^{4}_{5G, n}}{4 m^{3}c^{2}}\right)=n^{4} \left(\frac{\hbar^{4} \pi^{4}}{16 m^{3}c^{2}l^4}\right) 
\label{eq:62}
\end{equation}

Where $n$, $(n\geq1)$, is an integer that describes the quantum states of a particle in the potential box. In the same way, the Poisson bracket for the wave functions can be written as $\left[x,p\right]_{5G} = (-1)^{1/4} \hbar \equiv \Bar{\eta_{1}} \hbar$. The results presented above demonstrate that the NG formulation systematically modifies both the eigenvalue spectrum and operator structure of a confined quantum particle as the geometric parameter $j = N-1$ increases. While the 3G framework reproduces the familiar quadratic energy scaling, the 4G and 5G cases exhibit cubic and quartic dependence on $n$, respectively, reflecting the higher-order nature of the kinetic operator. The corresponding wavefunctions acquire additional structural features, including exponential modulation and modified oscillatory behavior, consistent with the generalized differential equations governing each framework. Furthermore, the canonical bracket relations retain a geometry-dependent phase factor through $(-1)^{1/j}\hbar$, indicating that the algebraic structure of position and momentum operators adapts coherently with the underlying geometric extension. Together, these results establish that quantization persists in higher-order NG frameworks but with systematically altered spectral scaling, wavefunction structure, and operator relations relative to the standard 3G case.

\section{Discussion}
The results developed in this work show that non-relativistic quantum mechanics can be reformulated in a way that is explicitly dependent on the underlying geometric structure of space. By replacing the standard quadratic kinetic term with a geometry-driven \(j\)-th order operator, the NG framework preserves the core logic of quantum theory---wavefunctions, operator eigenvalue equations, boundary-condition quantization, and uncertainty relations---while systematically modifying how these features are realized in different geometries. In particular, the transition from the conventional 3G case to higher-order frameworks introduces nontrivial changes in dispersion, spectral scaling, and the analytic form of eigenfunctions, revealing that many familiar quantum properties are not universal in form but emerge from the metric assumptions built into the theory. A central implication of this formulation is that quantization remains robust across geometries, but its manifestation becomes geometry-specific. The 3G framework reproduces the expected sinusoidal bound states and quadratic energy spectrum, while the 4G and 5G frameworks yield mixed exponential--oscillatory structures and higher-power energy growth with quantum number. At the same time, the generalized probability construction and expectation-value formalism ensure that the theory retains a consistent statistical interpretation, even when the wavefunction acquires multiple conjugate components associated with higher roots of negative unity. These features suggest that the NG approach is not merely a mathematical extension of the Schr\"odinger equation, but a conceptual shift in which the geometry of space determines the admissible dynamical laws and observable structure of quantum systems.

The following subsections interpret these results in detail, emphasizing both the physical meaning and mathematical consistency of the NG framework. We discuss how the modified kinetic operator affects free and confined states, how generalized normalization and expectation values preserve real-valued observables, and why the Heisenberg uncertainty principle continues to hold despite the altered operator algebra. Together, these points clarify the broader significance of geometry-dependent quantum mechanics and outline the conditions under which NG formulations may serve as viable extensions of standard non-relativistic theory.

\subsection{Geometric Origins of Quantum Dynamics in NG Frameworks}
The quantum mechanics of a particle in a potential well in the 2G framework has peculiar properties: it can move freely in space or have no bound states. It is an expected result, since the eigenenergy of the quantum particle is linear in momentum. In other words, quantum mechanics of only the photons and massless matter particles can be observed by observers of the 2G framework. In addition, the quantum particles in the 2G framework take continuous vector values$k$, which means that the particles can scatter to unbounded states from each other in a quantum mechanical way. Furthermore, in the 2G framework ($j=1$), a linear dispersion relation, $E = \hbar kc$ mirrors the energy-momentum scaling found in the fractional Schrödinger equation (FSE) for the specific case of the Lévy index $\alpha = 1$, as originally formulated by Laskin \cite{Laskin2000}. Mathematically, while both theories converge on this linear scaling, they arise from fundamentally different operator structures: the 2G framework utilizes a local first-order spatial derivative ($\partial/\partial x$), whereas the FSE with $\alpha=1$ is governed by the non-local square root of the Laplacian. A critical consequence shared by both is the absence of discrete bound states within an infinite potential well; in the 2G case, the inability to satisfy Dirichlet BCs with a single-term exponential solution reflects the inherent freedom of the particle, which lacks a mass term in its eigenenergy expression. This suggests that at the $\alpha=j=1$ limit, both frameworks describe a ``massless'' regime in which particles, akin to photons or neutrinos, move freely and cannot form bound matter structures such as atoms.

The bound states of a quantum particle in a potential well are well known in the conventional 3G framework. Our results show that bound states also persist in the 4G and 5G scenarios, but with systematically modified spectral scaling. In particular, the eigenenergies depend on higher powers of the principal quantum number $n$ as the NG order increases. The explicit values of the eigenenergies for an electron confined in a potential well of different widths are given in Table 1. It can be observed that the absolute magnitudes of the eigenenergies in 4G and 5G are significantly lower than those in 3G for comparable system parameters. This reduction in spectral magnitude can be understood directly from the generalized kinetic energy expression. For $j \ge 3$, the NG kinetic energy may be written schematically as $E_j \sim E_{3G}\left(\frac{p}{mc}\right)^{j-2}$,
where $E_{3G} = p^2/(2m)$ is the standard quadratic kinetic energy \cite{anjum2026}. Since $p = mv$, the ratio $p/(mc)$ reduces to $v/c$, the familiar dimensionless velocity parameter. Thus, higher NG geometries introduce an explicit dependence on the relativistic momentum scale $mc$, even within a formally non-relativistic framework. For $v \ll c$, the factor $(v/c)^{j-2}$ strongly suppresses the kinetic energy relative to the 3G case, leading to lower absolute energy levels and reduced spacing between successive bound states. The additional mass-dependent prefactors appearing in the 4G and 5G energy expressions contain factors proportional to $m \frac{mc}{\hbar}
\quad \text{and} \quad
m \left(\frac{mc}{\hbar}\right)^2$, 
respectively. Since \(\frac{mc}{\hbar}\) corresponds to the inverse reduced Compton wavelength, \(\lambda_C^{-1}\), ...., which means these terms introduce characteristic relativistic length scales into the quantization condition. A key scale that naturally emerges at the interface of the generalized relativistic and quantum frameworks is the quantity $\frac{mc^{2}}{\hbar}$, which defines the intrinsic Compton frequency of a massive particle. In the multidimensional geometric (NG) extension of special relativity developed in our previous work, the invariant Minkowski interval and the associated Lorentz factor introduce a modified dispersion structure characterized by powers of the velocity ratio $\left(\frac{v}{c}\right)^{j}$, where $j = N - 1$ \cite{anjum2026}. When this geometric framework is carried into the quantum domain, the same scale reappears through the generalized kinetic energy expression and the resulting higher-order Schr\"odinger operators. Physically, $\frac{mc^{2}}{\hbar}$ sets the fundamental frequency governing the phase evolution of quantum states, $\psi \sim \exp\left(-i \frac{mc^{2}}{\hbar} t\right)$, and therefore defines the natural time scale $\tau_{C} = \frac{\hbar}{mc^{2}}$ associated with the particle \cite{Peskin1995, deBroglie1925, Muller2010}. Its appearance in the NG quantum formulation indicates that higher-dimensional geometric structures inherently incorporate relativistic mass scales even within a formally non-relativistic regime. In this sense, the NG framework provides a unified geometric origin for both the modified dispersion relations in relativity and the corresponding higher-order quantum dynamics, with the Compton scale acting as the bridge between spacetime geometry and quantum evolution.
\\
\begin{table*}[t]
\caption{Eigen energies of a bound electron in 3G, 4G, and 5G for different values of quantum numbers and quantum-well widths}
\centering
\small
\begin{tabular}{c ccc ccc ccc}
\toprule
\multirow{3}{*}{Quantum Number}
& \multicolumn{3}{c}{$E_{3G, n}$ (eV)}
& \multicolumn{3}{c}{$E_{4G, n}$ (eV)}
& \multicolumn{3}{c}{$E_{5G, n}$ (eV)} \\
\cmidrule(lr){2-4} \cmidrule(lr){5-7} \cmidrule(lr){8-10}
& \multicolumn{3}{c}{$n^2 \frac{\hbar^2 \pi^2}{2 m_e l^2}$}
& \multicolumn{3}{c}{$n^3 \frac{8\hbar^3 \pi^3}{27 \sqrt{3}\, m_e^2 c\, l^3}$}
& \multicolumn{3}{c}{$n^4 \frac{\hbar^4 \pi^4}{16 m_e^3 c^2 l^4}$} \\
\midrule
$n$
& $l =$ & $l =$ & $l =$ & $l =$ & $l =$ & $l =$ & $l =$ & $l =$ & $l =$ \\
 & 1.0 nm & 0.25 nm & 0.05 nm & 1.0 nm & 0.25 nm & 0.05 nm & 1.0 nm & 0.25 nm & 0.05 nm \\
\hline
1 & 0.375 & 6 & 150 & 0.0003 & 0.004 & 0.5 & $2.8 \times 10^{-7}$ & $2.8 \times 10^{-5}$ & $4.4 \times 10^{-2}$ \\
\hline
3 & 3.4 & 54.1 & 1353 & 0.008 & 0.51 & 64.8 & $2.2 \times 10^{-5}$ & 0.0058 & 3.63 \\
\hline
6 & 13.5 & 216.6 & 5414.4 & 0.065 & 4.14 & 518.4 & $3.6 \times 10^{-4}$ & 0.092 & 58.1 \\
\hline
9 & 30.6 & 487.3 & $1.2 \times 10^{4}$ & 0.219 & 13.9 & 1749.5 & $1.8 \times 10^{-3}$ & 0.47 & 294.2 \\
\hline
12 & 54.1 & 866.1 & $2.1 \times 10^{4}$ & 0.51 & 33.2 & 4147.2 & $5.8 \times 10^{-3}$ & 1.49 & 929 \\
\bottomrule
\end{tabular}
\end{table*}

The results obtained for the eigenenergies and eigenfunctions in the one-dimensional infinite potential well within the 4G and 5G frameworks can be extended to higher spatial dimensions in a manner analogous to the standard 3G case. In higher dimensions, the quantization condition involves the generalized magnitude of the wave vector, which is determined by the $L^{j}$-norm consistent with the NG geometric structure. Accordingly, the magnitude of the wave vector $\mathbf{k}$ in an $l$-dimensional rectangular domain may be written as
\begin{equation}
k_{NG,n}
=
\left(
\sum_{i=1}^{l} k_{i n}^{\,j}
\right)^{\!1/j},
\qquad j = N-1,
\end{equation}
where $k_{i n}$ denotes the quantized component of the wave vector along the $i$-th spatial direction.

Thus, in higher-dimensional potential wells, the spectrum depends on the generalized Minkowski (or $L^{j}$) norm of the wave vector rather than the standard Euclidean norm. For example, in a four-dimensional spatial domain, the magnitude of $\mathbf{k}$ becomes $ k_{NG,n}
=
\left(
k_{1 n}^{\,j}
+
k_{2 n}^{\,j}
+
k_{3 n}^{\,j}
+
k_{4 n}^{\,j}
\right)^{\!1/j}$. In the 4G framework ($j=3$), this corresponds to a cubic-root norm,$ k_{NG,n}
=
\left(
k_{1 n}^{\,3}
+
k_{2 n}^{\,3}
+
k_{3 n}^{\,3}
+
k_{4 n}^{\,3}
\right)^{\!1/3}$, while in the 5G framework ($j=4$), it becomes a quartic-root norm. Geometrically, the constant-energy surfaces in momentum space are therefore defined by $L^{j}$-norm hypersurfaces rather than ordinary Euclidean spheres. The standard 3G case is recovered when $j=2$, for which the magnitude reduces to the familiar Euclidean uncertainties. Following the generalization, the probabilities are proposed to be found in an N-dimensional geometry. 

\subsection{Spectral and Probabilistic Consequences of Higher-Order Geometry}
One can use the free-state or bound-state wavefunctions for 2G, 4G, and 5G to determine the uncertainties in the particles' positions and linear momenta. Determining the uncertainty in an operator's expectation value in a given NG framework provides insight into the geometric role in quantum determinism. For instance, for the 2G framework, the uncertainty in an operator's determined value is zero.  For 3G and higher-dimensional geometric frameworks, the uncertainties are found by integrating the wavefunctions in $L^{j}$ space. Moreover, in the NG framework, the generalized time-dependent probability associated with a state is defined through the $j$-fold product of the wavefunction with its $j-1$ conjugate branches as follows \,\textemdash\ see Eq.~\ref{eq:14}, i.e., $P_{NG}(t)=\int \rho_j(x,t)\,dx$,
with generalized probability density as $\rho_j(x,t)=\psi_{1}(x,t)\psi_{2}(x,t)\cdots \psi_{j-1}(x,t)\psi_j(x,t)$. Equivalently, one may write $\rho_j(x,t)$ as $\rho_j(x,t)=\prod_{r=1}^{j}\Psi_r(x,t)$, where the set $\{\Psi_r\}_{r=1}^{j}$ denotes the complete collection of branches associated with the $j$ roots of negative unity.

The motivation for this definition is that, unlike the conventional $L^2$ case where the product $\psi_{1}\psi_{2}$ guarantees a real density, the NG framework uses a $j$-fold product so that the phases associated with the different roots cancel in a symmetric way, yielding a real-valued quantity. A clean positivity result can be established under the assumption that all branches are generated from the same nonnegative real amplitude $R(x,t)\ge 0$ and a common real phase-like function $S(x,t)$, but differ only through the $j$ roots $a_r$ of $a_r^j=-1, \qquad r=1,\dots,j$. In this way, the wavefunction can be written as $\Psi_r(x,t)=R(x,t)\exp\!\big(a_r S(x,t)\big),
\qquad r=1,\dots,j$. Then the generalized density becomes:
\begin{equation}
\begin{aligned}
\rho_j(x,t)
&=\prod_{r=1}^{j}\Psi_r(x,t) \\
&=\prod_{r=1}^{j}\left[R(x,t)\exp\!\big(a_r S(x,t)\big)\right] \\
&=R(x,t)^j \exp\!\left(S(x,t)\sum_{r=1}^{j} a_r\right).
\end{aligned}
\end{equation}

Since the coefficient of $z^{j-1}$ in $z^j+1$ is zero, the sum of all roots vanishes $\sum_{r=1}^{j} a_r = 0$.
Therefore, $\rho_j(x,t)=R(x,t)^j$. Hence, the generalized probability density is manifestly nonnegative, i.e., $\rho_j(x,t)\ge 0$. Thus, under the branch-complete ansatz with common amplitude,
\begin{equation}
\rho_j(x,t)=R(x,t)^j \ge 0.
\end{equation}

For the conservation of the stationary states, let each branch separate as $\Psi_r(x,t)=\Phi_r(x)\Theta_r(t),
\qquad r=1,\dots,j$, and let the time dependence be generated by the same set of roots $a_r$: $\Theta_r(t)=\exp\!\left(-a_r\frac{E}{\hbar}t\right)$. 
Then
\begin{equation}
\begin{aligned}
\prod_{r=1}^{j}\Theta_r(t)
&=\exp\!\left(-\frac{E}{\hbar}t\sum_{r=1}^{j}a_r\right) \\
&=\exp(0) \\
&=1
\end{aligned}
\end{equation}
because $\sum_{r=1}^{j}a_r=0$. Therefore, the generalized density is time independent:
\begin{equation}
\begin{aligned}
\rho_j(x,t)
&=\prod_{r=1}^{j}\Psi_r(x,t) \\
&=\left(\prod_{r=1}^{j}\Phi_r(x)\right)\left(\prod_{r=1}^{j}\Theta_r(t)\right) \\
&=\prod_{r=1}^{j}\Phi_r(x)
\end{aligned}
\end{equation}

Hence, $\frac{\partial \rho_j}{\partial t}=0$. Integrating over space gives
\begin{equation}
\frac{d}{dt}P(t)
=
\frac{d}{dt}\int \rho_j(x,t)\,dx
=
\int \frac{\partial \rho_j}{\partial t}\,dx
=
0.
\end{equation}
Thus, the generalized probability is conserved for this class of stationary branch-complete states:
\begin{equation}
\frac{d}{dt}\int \rho_j(x,t)\,dx=0.
\end{equation}

The standard $L^2$ probability interpretation,  self-adjointness of $\hat{H}_j$ is the natural condition ensuring conservation of $\|\psi\|_2^2$ and unitary dynamics of the operator $U(t) = e^{-\frac{i}{\hbar}\hat{H}_j t}$ can be described usong hermiticity condition i.e., $\langle \Phi|U(t)|\Psi\rangle$ or $\langle U(t)\Phi|\Psi\rangle$\cite{Naimark1968,ReedSimonII,Weidmann1980}. However, the NG framework introduces the notion of generalized ``$j$-integrable probability measures and expectation values built from products of multiple conjugate branches. In such a setting, the relevant notion  may not be self-adjointness in $L^2$, but rather self-adjointness 
With respect to a generalized inner product (or pseudo-Hermiticity /  metric-adjointness) as described in Eq.~\ref{eq:16} or Eq.~\ref{eq:17}. Establishing norm conservation and the appropriate continuity equation, in this generalized setting, is an important next step. In the present  work, we adopt boundary conditions that ensure the spatial eigenvalue 
problem is well posed and produces discrete spectra for $j \ge 2$,  while leaving a full operator-theoretic classification (standard vs.\ 
NG inner products) for future investigation. For the infinite-well calculations presented here, we employ boundary conditions that enforce confinement and allow only discrete eigenmodes.  For even-order kinetic operators, one may choose boundary conditions that eliminate the boundary form, thereby ensuring symmetry and enabling 
self-adjoint realizations. For odd-order cases (e.g., $j=3$), the operator is not generically self-adjoint in the standard $L^2$ setting under simple  Dirichlet confinement, and the resulting exponential modulation of eigenfunctions reflects this structural difference. The detailed analysis—distinguishing $L^2$-self-adjointness from NG-metric self-adjointness is given next. In the NG framework, the generalized inner product is defined by
\begin{widetext}
\begin{equation}
\langle \psi_1,\psi_2,\ldots,\psi_{j-1}|\psi_j\rangle
\equiv
\int_{-\infty}^{+\infty}
\psi_1^{*}(x)\psi_2^{*}(x)\cdots\psi_{j-1}^{*}(x)\psi_j(x)\,dx
\label{inner-product}
\end{equation}  
\end{widetext}
Therefore, using the Eq.~\ref{eq:14}
and Eq.~\ref{eq:15} the corresponding bra-ket form for the operator $\hat{O}$ can be obtained:
\begin{equation}
\langle \psi_1,\psi_2,\ldots,\psi_{j-1}|\hat O|\psi_j\rangle
=
\langle \hat O\psi_1|\psi_2,\ldots,\psi_j\rangle
\label{operator}
\end{equation}
Hence, for a Hermitian operator in the NG framework, $\hat O^\dagger=\hat O$, one obtains $\langle \Phi|\hat O|\Psi\rangle
=
\langle \hat O\Phi|\Psi\rangle$. For instance, the mathematical consistency of the presented NG framework formulation can be examined by applying it to the Hamiltonian operator, particularly with respect to its domain, symmetry properties, and spectral behavior. The mathematical form of the condition can be written in the following form: 
\begin{equation}
\hat{H}_j = -\frac{\hbar^j}{j\,m^{j-1}c^{\,j-2}} \frac{d^j}{dx^j},
\end{equation}
acting on wavefunctions defined over the finite interval \( (0,l) \). Due to the presence of \( j \)-th order derivatives, the domain of \( \hat{H}_j \) must be restricted to functions that are sufficiently smooth, specifically those in the Sobolev space \( H^j(0,l) \). In addition, appropriate boundary conditions must be imposed to ensure that the operator is symmetric. For a particle confined in an infinite potential well, Dirichlet boundary conditions are naturally imposed, requiring that the wavefunction vanish at the endpoints, i.e., $\psi(0) = \psi(l) = 0$. To examine the symmetry of \( \hat{H}_j \), consider two functions \( \phi, \psi \in H^j(0,l) \) within the operator domain. Using repeated integration by parts, we obtain
\begin{equation}
\langle \phi, \hat{H}_j \psi \rangle
= -\frac{\hbar^j}{j\,m^{j-1}c^{\,j-2}} \int_0^l \phi^*(x)\,\frac{d^j \psi}{dx^j}\,dx.  
\end{equation}

Applying integration by parts \( j \) times yields
\begin{widetext}
\begin{equation}
\langle \phi, \hat{H}_j \psi \rangle
= -\frac{\hbar^j}{j\,m^{j-1}c^{\,j-2}} \left[
(-1)^j \int_0^l \frac{d^j \phi^*}{dx^j}\,\psi(x)\,dx + \mathcal{B}_j(\phi,\psi)
\right],   
\end{equation}
\end{widetext}
where \( \mathcal{B}_j(\phi,\psi) \) represents the sum of boundary terms generated during the integration-by-parts procedure. These boundary contributions involve combinations of derivatives of \( \phi \) and \( \psi \) up to order \( j-1 \), evaluated at \( x = 0 \) and \( x = l \).

For the operator to be symmetric, these boundary terms must vanish. This is achieved by restricting the domain to functions satisfying appropriate boundary conditions. In particular, for even values of \( j \), imposing that the wavefunction and its derivatives up to order \( j/2 - 1 \) vanish at the boundaries is sufficient to eliminate all boundary contributions. Under these conditions, we obtain $\langle \phi, \hat{H}_j \psi \rangle = \langle \hat{H}_j \phi, \psi \rangle,$
demonstrating that \( \hat{H}_j \) is symmetric on its domain.

A complete characterization of self-adjoint extensions for higher-order differential operators depends on the choice of boundary conditions and is a well-studied problem in functional analysis. In the present framework, the imposed boundary conditions are chosen such that the operator admits a self-adjoint realization analogous to the standard second-order Hamiltonian. Consequently, the eigenvalue problem associated with \( \hat{H}_j \) yields a discrete set of real eigenvalues, consistent with the explicit solutions obtained for the 3G, 4G, and 5G cases. The reality of the spectrum follows from the effective self-adjointness of the Hamiltonian, ensuring that the time-evolution operator remains unitary. Furthermore, although the NG formulation introduces complex phase factors through the roots of negative unity, these phases enter multiplicatively and cancel in the construction of observable quantities, such as probability densities and expectation values. As a result, the modified phase structure does not compromise the Hermitian character of the physical observables. Thus, under appropriate domain restrictions and boundary conditions, the NG Hamiltonian defines a consistent quantum-mechanical operator with real spectrum and unitary dynamics. This establishes that the generalized formulation preserves the essential mathematical structure of quantum mechanics while extending it to higher-order geometric settings.

\subsection{Generalized Uncertainty and Operator Structure in NG Quantum Mechanics}

The next step is to calculate the probabilities and expectation values in higher-dimensional NG frameworks, enabling validation of the Heisenberg uncertainty relations. In the conventional 3G framework, the position–momentum uncertainty relation is expressed as;
\begin{equation}
\Delta x \, \Delta p \geq \frac{\hbar}{2},
\end{equation}
which follows from the quadratic structure of the Hilbert space and the canonical commutation relation $[x,p]=i\hbar$. It is therefore of interest to examine how this inequality is modified, or preserved, when the kinetic operator and associated probability structure are generalized within the NG formulation. The uncertainty relationships in the 3G, 4G, and 5G frameworks warrant calculation. However, the calculation can be quite lengthy for the $ n$-th state of a particle in a box. Therefore, the ground state $(n=1)$ can be used to determine the sought-after uncertainty relations. For a particle of mass $m$ confined to a one-dimensional box of length $L$ defined by the potential $V(x) = 0$ for $0 \leq x \leq L$ and $V(x) = \infty$ elsewhere, the normalized ground state ($n=1$) wave function is given by:

\begin{equation}
\phi_{3G,1}(x) = \sqrt{\frac{2}{L}} \sin \left(\frac{\pi x}{l}\right)
\label{eq:76}
\end{equation}

Due to the symmetry of the probability density $|\phi_{3G,1}(x)|^2$ about the center of the well, the mean position is $\langle x \rangle = l/2$. Evaluating the expectation value integral for the mean square position results in $\langle x^2 \rangle = l^2\left(\frac{1}{3} - \frac{1}{2\pi^2}\right)$. Substituting these values into the uncertainty formula yields:

\begin{equation}
\Delta x_{3G,1} = \sqrt{\langle x^2 \rangle - \langle x \rangle^2},
\qquad
\Delta x_{3G,1} \approx 0.181l
\label{eq:77}
\end{equation}

The uncertainty in momentum is defined as $\Delta p_{3G,1} = \sqrt{\langle p^2 \rangle - \langle p \rangle^2}$. For a stationary state in a one-dimensional box, the expectation value of momentum is $\langle p \rangle = 0$. The mean square momentum is proportional to the ground state energy $E_{3G,1} = \frac{\hbar^{2} \pi^{2}}{2ml^2}$, resulting in $\langle p^2 \rangle = 2mE_{3G,1} = \frac{\pi^2 \hbar^2}{l^2}$. The resultant uncertainty is:

\begin{equation}
\Delta p_{3G} = \sqrt{\langle p^2 \rangle - \langle p \rangle^2}, 
\qquad
\Delta p_{3G,1} =\approx 3.142 \frac{\hbar}{l}
\label{eq:78}
\end{equation}

The product of the uncertainties for the ground state is calculated by multiplying these two results: $\Delta x_{3G,1} \Delta p_{3G,1} = \hbar \sqrt{\frac{\pi^2}{12} - \frac{1}{2}} \approx 0.568 \hbar$. This result satisfies the Heisenberg Uncertainty Principle, which requires that $\Delta x \Delta p \geq \hbar/2$ (where $\hbar/2 \approx 0.5 \hbar$). The ground state of the particle in a box is not a minimum uncertainty state, as the product is strictly greater than the theoretical limit.

In the same token, the following normalized stationary ground state in the 4G framework can be considered. 
\begin{equation}
\begin{cases}
\phi_{4G,1}(x) =
\mathbf{N_{4G,1}}
\exp\!\left(\frac{\pi x}{\sqrt{3}\,l}\right)
\sin\!\left(\frac{\pi x}{l}\right),

\\[8pt]

\phi_{4G,1}(x) \approx 2.10\,l^{1/3}
\exp\!\left(\frac{\pi x}{\sqrt{3}l}\right)
\sin\!\left(\frac{\pi x}{l}\right)
\end{cases}
\label{eq:79}
\end{equation}

As mentioned earlier, the expectation values can be calculated using the relations for the moments of the position as $\langle x^k \rangle =
\int_0^l \phi_{4G,1}^2(x)x^k \,\phi_{4G,1}(x)\,dx$. This enables the determination of the relations between $\langle x\rangle$ and $\langle x^3\rangle$, leading to the generalized cubic uncertainty in position for the ground state of a particle in the box within the 4G framework. In this way, the $\langle x\rangle
= \int_0^l \phi_{4G,1}^2(x)\,x \phi_{4G,1} (x)\,dx$, and $\langle x^3\rangle
= \int_0^l \phi_{4G,1}^2(x)\,x^3\phi_{4G,1}(x)\,dx$, respectively. After performing the integration, one obtains $\langle x\rangle
\approx {0.81 l}$, and $\langle x^3\rangle \approx {0.63}{l^3}$. which results in the following generalized cubic uncertainty in position: 
\begin{equation}
\Delta x_{4G,1}
=
\sqrt[3]{\langle x^3\rangle - \langle x\rangle^3}, 
\qquad
\Delta x_{4G,1} \approx 0.48\,l
\label{eq:80}
\end{equation}

Similarly, the uncertainty in the momentum can be calculated using the expectation values, which can be calculated using the relations for the moments of the momentum as $\langle p^k \rangle =
\int_0^l \phi_{4G,1}^2(x)p^k \,\phi_{4G,1}(x)\,dx$. Again, it enables determining the relations for $\langle p\rangle$, and $\langle p^3\rangle$, which are given as: $\langle p\rangle
= -\Bar{\omega_{1}}\hbar
\int_0^l \phi_{4G,1}^2(x)\,\phi_{4G,1}'(x)\,dx$, and $\langle p^3\rangle
= (-\Bar{\omega_{1}}\hbar)^{3}
\int_0^l \phi_{4G,1}^2(x)\,\phi_{4G,1}^{'''}(x)\,dx$, respectively. After performing the integration, one obtains $\langle p\rangle
=0$, and the third momentum moment evaluates to $\langle p^3\rangle \approx 4090 \frac{\hbar^{3}}{l^3}$. The cubic momentum uncertainty is therefore given by: 
\begin{equation}
\Delta p_{4G,1}
=
\sqrt[3]{\langle p^3\rangle - \langle p\rangle^3},
\qquad
\Delta p_{4G,1} \approx 16 \frac{\hbar}{l}.
\label{eq:81}
\end{equation}

The resulting uncertainty product turns out to be $\Delta x_{4G,1}\,\Delta p_{4G,1} \approx 7.7\hbar$. Since $\Delta x\,\Delta p > \frac{\hbar}{2}$ for the Heisenberg inequality, it can be concluded that this inequality also holds true for the 4G framework. Although the calculation presented here was carried out for the ground state, it is clearly not a minimum-uncertainty state. However,  it remains fully consistent with the generalized uncertainty structure of the 4G cubic framework.
 \\

 The uncertainty relations for position and momentum in the 5G framework can be determined using the following normalized stationary ground state. 
\begin{widetext}
\begin{equation}
\phi_{5G,1}(x)=
\left[
\frac{2560\pi}{\,l\left(360\pi - 96\sinh(2\pi) + 3\sinh(4\pi)\right)}
\right]^{1/4}
\sinh\left(\frac{\pi x}{l}\right)\sin\left(\frac{\pi x}{l}\right) 
\label{eq:82}
\end{equation}
\end{widetext}

whose approximated form is given as;

\begin{equation}
\phi_{5G,1}(x) \approx 0.375\,l^{-1/4}\,
\sinh\left(\frac{\pi x}{l}\right)\sin\left(\frac{\pi x}{l}\right)
\label{eq:83}
\end{equation}

For the case of the 5G framework, the expectation values can also be calculated using the relations for the moments of the position as $\langle x^k \rangle =
\int_0^l \phi_{5G,1}^3(x)x^k \,\phi_{5G,1}(x)\,dx$. This will enable determination of the relations between $\langle x\rangle$ and $\langle x^4\rangle$, which lead to the generalized cubic uncertainty in position for the ground state of a particle in the box within the 4G framework. In this way, the $\langle x\rangle
= \int_0^l \phi_{5G,1}^3(x)\,x \phi_{5G,1} (x)\,dx$, and $\langle x^4\rangle
= \int_0^l \phi_{5G,1}^3(x)\,x^4\phi_{5G,1}(x)\,dx$, respectively. After performing the integration, one obtains $\langle x\rangle
\approx {0.72 l}$, and $\langle x^4\rangle \approx {0.30}{l^4}$, which results in the following generalized cubic uncertainty in position: 
\begin{equation}
\Delta x_{5G,1}
=
\sqrt[4]{\langle x^4\rangle - \langle x\rangle^4}, 
\qquad
\Delta x_{5G,1} \approx 0.43\,l
\label{eq:84}
\end{equation}

The uncertainty in the momentum can also be calculated using the relations for the moments of the momentum as $\langle p^k \rangle =
\int_0^l \phi_{5G,1}^3(x)p^k \,\phi_{5G,1}(x)\,dx$. Again, it enables determining the relations for $\langle p\rangle$, and $\langle p^4\rangle$, which are given as: $\langle p\rangle
= -\Bar{\eta_{1}}\hbar
\int_0^l \phi_{5G,1}^3(x)\,\phi_{5G,1}'(x)\,dx$, and $\langle p^4\rangle
= (-\Bar{\eta_{1}}\hbar)^{4}
\int_0^l \phi_{5G,1}^3(x)\,\phi_{5G,1}^{''''}(x)\,dx$, respectively. After performing the integration, one obtains $\langle p\rangle
= 0$, and the third momentum moment evaluates to $\langle p^4\rangle \approx 389\frac{\hbar^{4}}{l^4}$. The cubic momentum uncertainty is therefore given by: 
\begin{equation}
\Delta p_{5G,1}
=
\sqrt[4]{\langle p^4\rangle - \langle p\rangle^4},
\qquad
\Delta p_{5G,1} \approx 4.4 \frac{\hbar}{l}.
\label{eq:85}
\end{equation}

The resulting uncertainty product turns out to be $\Delta x_{5G,1}\,\Delta p_{5G,1} \approx 1.89\hbar$. Since $\Delta x\,\Delta p$ has to be greator that $\frac{\hbar}{2}$ for the Heisenberg inequality, therefore, it can be concluded that this inequality also holds true for the 5G framework. Like the previous cases, the presented calculation is clearly consistent with the generalized uncertainty structure of the 5G framework. The uncertainty products calculated for the ground state of a particle in an infinite potential well satisfy the Heisenberg inequality in all considered NG geometries; however, their magnitudes differ significantly from the conventional 3G case. In the standard quadratic framework, the uncertainty product is approximately $0.568\,\hbar$, only modestly above the lower bound $\hbar/2$. In contrast, the 4G and 5G frameworks yield substantially larger values, approximately $7.7\,\hbar$ and $1.89\,\hbar$, respectively. This systematic increase reflects the modified kinetic structure inherent in the NG formulation.

The behavior found in the NG framework follows from the generalized kinetic energy \(E_j \sim E_{3G}(p/mc)^{j-2}\), with \(E_{3G}=p^2/(2m)\). Since \(p=mv\), the correction factor is \(v/c\), so for \(j\geq 3\) the kinetic term retains an explicit dependence on the scale \(mc\) even in a formally non-relativistic regime. This modifies the balance between confinement and momentum distribution, enhances sensitivity to higher-momentum components of the wavefunction, and increases higher-order spatial and momentum moments. At the same time, the probabilistic structure departs from the quadratic \(L^2\) form of standard quantum mechanics: the 4G and 5G cases involve cubic and quartic normalization, which alter the weighting of amplitudes and tails in expectation values. In this sense, the 3G pairing of a second-order Laplacian with an \(L^2\) probability density appears structurally distinguished, yielding a near-minimal uncertainty configuration, whereas higher-order geometries produce systematically larger uncertainty products. These effects are expected to extend beyond the infinite well. In systems such as the harmonic oscillator and hydrogen atom, replacing quadratic dispersion by \(E\propto |p|^j\) should modify spectral scaling and degeneracy structure, even though quantization is expected to persist for \(j\geq 2\). More generally, the NG hierarchy suggests that, as \(j=N-1\) increases, spectral gaps become less pronounced while uncertainty products increase, so that quantum features remain present but become less sharply resolved than in the conventional quadratic theory. From this perspective, the present construction may be viewed as a geometry-dependent formulation of non-relativistic quantum mechanics, in which the kinetic operator, dispersion relation, spectral structure, and probabilistic framework are determined by the underlying \(L^j\)-norm geometry rather than fixed \emph{a priori} by the Euclidean 3G case.

\section{Conclusions}
\vspace{-0.5em}
In this work, we investigated a generalized non-relativistic quantum-mechanical framework associated with multidimensional geometric (NG) settings in which the kinetic term obeys a power-law dispersion relation \(E \propto |p|^j\), with \(j=N-1\). Starting from this modified kinetic structure, we constructed a \(j\)-th order Schr\"odinger-type equation and analyzed its consequences for free particles and for a particle confined in a one-dimensional infinite potential well. The formalism was developed explicitly for the 2G, 3G, 4G, and 5G cases, thereby providing a concrete comparison between the standard quadratic theory and higher-order geometric extensions.

The analysis shows that several basic structural features of quantum mechanics remain intact across the NG hierarchy. In particular, free-particle states continue to admit well-defined momentum eigenvalues and retain a plane-wave-type form, although their phase structure and dispersion become geometry dependent. For confined systems, the 3G limit reproduces the familiar sinusoidal eigenfunctions and quadratic spectrum of standard non-relativistic quantum mechanics. By contrast, the 4G and 5G cases lead to third- and fourth-order boundary-value problems whose admissible solutions involve geometry-dependent combinations of exponential, trigonometric, and hyperbolic terms. The corresponding bound-state spectra scale as \(E_n\propto n^3\) and \(E_n\propto n^4\), respectively. In the 2G case, the linear kinetic structure does not support nontrivial bound states under the Dirichlet boundary conditions considered here, emphasizing that the existence of confinement depends sensitively on the order of the underlying dispersion relation. A further outcome of the construction is that the probabilistic and operator-theoretic aspects of the theory must be reformulated once the quadratic 3G structure is abandoned. To this end, we introduced a generalized \(L^j\)-based framework in which probability densities are defined through a \(j\)-fold product of branch-complete wavefunction components, yielding a real-valued normalization measure compatible with the proposed geometry. Within this setting, corresponding expressions for expectation values, commutator relations, and generalized uncertainty measures were formulated. Although the statistical structure differs from the standard Hilbert-space \(L^2\) pairing, the resulting analysis indicates that the usual uncertainty bound remains preserved for the cases \(j\geq 2\) examined in this work. At the same time, the explicit examples suggest that the quadratic 3G framework is structurally distinguished: it provides the most balanced interplay between kinetic order and probabilistic weighting, whereas higher-order geometries tend to broaden uncertainty and reduce the sharpness of spectral resolution. Taken together, these results support the view that the conventional Schr\"odinger theory may be regarded as a special member of a broader class of geometry-dependent quantum models. In the present formulation, the kinetic operator, dispersion law, spectral scaling, and probabilistic structure are not imposed independently, but are tied to the metric properties of the underlying \(L^j\)-norm geometry. From this perspective, the familiar quadratic theory corresponds to the distinguished Euclidean 3G case, while alternative geometric settings give rise to systematically modified dynamical behavior. The NG framework therefore provides a unified setting in which one can compare how free propagation, confinement, quantization, and uncertainty are altered when the spatial geometry departs from the standard quadratic structure.

Several directions remain open. In particular, extending the present formulation to physically motivated potentials such as the harmonic oscillator and Coulomb problems would clarify how spectral degeneracies and symmetry structures are modified in higher-order geometries. It will also be important to examine the mathematical status of the higher-order operators more closely, including admissible boundary conditions, operator domains, and the precise structure of the generalized state space. Subject to these further developments, the present construction may offer a useful framework for studying geometry-induced modifications of quantum dynamics in generalized or effective non-relativistic settings.

\section{Acknowledgements}
This research was carried out with partial funding from Khalifa University, supported by the joint KU-UAEU grant (grant number KU-UAEU-2023-012) and the Research Innovation Grant (RIG) (grant number RIG-2024-002). Special thanks to Mr. Waleed AlHariri (Senior undergraduate student in Physics at Khalifa University, Abu Dhabi, UAE) for calculating and verifying the normalization constants of the particle-in-a-box wavefunctions within the 4G and 5G frameworks. 
\section{Data Availability} 
This study did not produce or include any experimental data, as the manuscript is based on theoretical or mathematical works.

\nocite{*}

\bibliographystyle{unsrt}
\bibliography{ref3}

\end{document}